\documentclass[fleqn,usenatbib]{mnras}

\usepackage{newtxtext,newtxmath}
\usepackage{stfloats}
\usepackage[flushleft]{threeparttable}
\usepackage{subfigure}
\usepackage[T1]{fontenc}
\usepackage{setspace}

\usepackage{graphicx}	
\usepackage{amsmath}	

\usepackage{amssymb}	
\usepackage[justification=centering]{caption}
\usepackage{float}
\usepackage{longtable}
\usepackage{supertabular}

\title[Periodic X-ray Sources in 47 Tuc]{Periodic X-ray sources in the Massive Globular Cluster 47 Tucanae: Evidence for Dynamically Formed Cataclysmic Variables }

\author[Bao, Li \& Cheng]{
Tong Bao$^{1,2}$\thanks{E-mail: baotong@smail.nju.edu.cn}, 
Zhiyuan Li$^{1,2}$\thanks{E-mail: lizy@nju.edu.cn}, Zhongqun Cheng$^{3,4}$\thanks{E-mail: chengzq@whu.edu.cn}
\\
$^{1}$School of Astronomy and Space Science, Nanjing University, Nanjing 210046, China\\
$^{2}$Key Laboratory of Modern Astronomy and Astrophysics (Nanjing University), Ministry of Education, Nanjing 210046, China\\
$^{3}$School of Physics and Technology, Wuhan University, Wuhan 430072, China\\
$^{4}$2WHU–NAOC Joint Center for Astronomy, Wuhan University, Wuhan 430072, China
}

\date{Accepted XXX. Received YYY; in original form ZZZ}

\pubyear{2022}

\begin{document}
\maketitle
\begin{abstract}
We present a systematic study of periodic X-ray sources in the massive globular cluster 47 Tuc, utilizing deep archival {\it Chandra} observations that resolve the cluster core and recently available {\it eROSITA} observations that cover the cluster outskirt. By applying the Gregory-Loredo algorithm, we detect 20 periodic signals among 18 X-ray sources, ranging between 205--95731 second.
Fourteen periods are newly discovered in the X-ray band. 
We classify these periodic sources into four quiescent low-mass X-ray binaries, one milli-second pulsar, two coronally-active binaries and eleven cataclysmic variables (CVs), based on their X-ray temporal and spectral properties, as well as multi-band information. 
Despite a small sample subject to potential selection bias against faint and non-magnetic CVs, the 11 CVs together define an orbital period distribution significantly different from that of the CVs previously found in the solar neighborhood and the Galactic bulge. 
In particular, there exists in 47 Tuc an apparent paucity of short-period CVs below the period gap, which might be attributed to a high occupation fraction of non-magnetic CVs. 
Also characteristic of the 47 Tuc CVs are an overabundance of long-period CVs with a subgiant donor, a
substantial fraction of CVs within the period gap, and a steep radial surface density profile.
These are best understood as a group of CVs having recently formed via dynamical interactions in the dense cluster core. 
Despite sufficient sensitivity of the X-ray data, only one periodic source is found between one-third of the half-light radius and the tidal radius, the nature of which is unclear.

\end{abstract}



\defcitealias{2003ApJ...596.1177E}{E03a}
\defcitealias{2003ApJ...596.1197E}{E03b}
\defcitealias{2018MNRAS.475.4841R}{RS18}

\section{Introduction}
As self-gravitationally bound systems evolving with a high stellar density, globular clusters (GCs) have long been regarded as the {\it mecca} for 
close binaries, in particular, low-mass X-ray binaries (LMXBs) with an accreting black hole (BH) or neutron star (NS), cataclysmic variables (CVs) with an accreting white dwarf (WD), coronally active binaries (ABs), as well as their potential descendants, e.g., milli-second pulsars (MSPs) and blue straggler stars. 
Theoretically, the formation and evolution of these exotic objects are strongly affected by or directly resulted from stellar dynamical encounters (including flybys, tidal captures and collisions) in the dense GC environment.
Moreover, close binaries are more likely to participate in dynamical encounters than single stars, during which they can release (or sometimes absorb) a substantial amount of kinetic energy \citep{1975AJ.....80..809H,1975MNRAS.173..729H},  
thereby playing a crucial role in the gravothermal evolution of the host cluster \citep{2003gmbp.book.....H}.
The demography of close binaries in GCs is key to our understanding of not only the profound physics of binary formation and evolution, but also the long-term evolution of GCs and their potential role as the factory of gravitational waves and seeds of massive BHs. 

Compared to normal stars, close binaries are significantly more luminous at certain wavelengths and thus are more easily discerned even in the dense cluster core, making themselves unique probes for the fundamental dynamical processes in which they are frequently involved \citep{2001Sci...292.2290G,2003ApJ...588..452H,2003ApJ...596.1177E,2006ApJ...651.1098H}.
In the X-ray band, while the presence and overabundance of luminous LMXBs (with X-ray luminosities  $L_{\rm X} \gtrsim10^{35}\rm~erg~s^{-1}$) in GCs were already recognized since the {\it Uhuru} era \citep{1975ApJ...199..307K,1975ApJ...199L.143C},
it is the {\it Chandra X-ray Observatory} that finally opened up the possibility of resolving weak X-ray sources (with $10^{30}{\rm~erg~s}^{-1} \lesssim L_{\rm X} \lesssim 10^{34}\rm~erg~s^{-1}$) in essentially every Galactic GC \citep{2008A&A...488..921B,2008A&A...490..641S,2009ApJ...705..175L,2010ApJ...712..380L,2009ApJ...697..224H,2010AIPC.1314..135H}. This effectively makes available a sizable sample of weak X-ray sources, ranging from a few to a few hundred sources per cluster, for statistical analyses and population studies \citep{2002ApJ...569..405P,2005ApJ...625..796H,2006ApJ...646L.143P,2012ApJ...756..147M}. 

Indeed, statistical studies of the weak X-ray sources have been fruitful. 
In particular, \citet{2003ApJ...591L.131P} and \citet{2006ApJ...646L.143P} found a positive correlation betwen the number of detected CVs and the stellar encounter rate of individual GCs, which led them to suggest that the majority of CVs in GCs have a dynamical origin, similar to the well-established case of LMXBs.
On the other hand, \citet{2018ApJ...858...33C}, who used the cumulative X-ray emissivity (with luminous LMXBs subtracted) as a proxy of the abundance of weak X-ray sources, found an underabundance rather than overabundance of CVs and ABs in most GCs relative to the Galactic field (see also \citealp{2020MNRAS.492.5684H}).
This was understood as stellar encounters being efficient in disrupting a large fraction of primordial, wide binaries before they can otherwise evolve into CVs and ABs \citep{2018ApJ...858...33C}. 
This finding is supported by the MOCCA simulations \citep{2016MNRAS.462.2950B,2017MNRAS.464.4077B}, which predicted that detectable CVs in GCs are predominantly composed of CVs formed via the common envelope phase rather than dynamical interactions.
This holds in the latest version of MOCCA simulation \citep{2019MNRAS.483..315B}, which, on the other hand, also predicts that strong dynamical interactions are able to trigger CV formation in binaries that otherwise would not have become CVs.
Direct observational evidence for dynamically-formed CVs in GCs, however, remains scarce. 

The spatial (radial) distribution of the weak X-ray sources also offers important insight to dynamical processes unique to the GC environment, although a meaningful analysis of this kind is necessarily restricted to those GCs with a sufficiently large number of detected X-ray sources. In a series of studies, Cheng and collaborators \citep{2019ApJ...876...59C,2019ApJ...883...90C,2020ApJ...892...16C,2020ApJ...904..198C} examined the radial surface density profile of weak X-ray sources in 47 Tuc, Terzan 5, M28 and $\omega$ Cen. They found a significant dip in the radial profile, which can be understood as the effect of mass segregation, i.e., the
X-ray sources as close binaries are on average heavier than the single
stars and are more likely to sink into the cluster core through two-body relaxation.

Despite the aforementioned observational and theoretical advances in our understanding of the GC X-ray populations, key questions regarding their origin, evolutionary route and present-day abundances, remain largely unanswered. 
In this and subsequent work, we attempt to shed light on the demography of the GC X-ray sources, in particular CVs, by examining their temporal properties, taking advantage of the extensive {\it Chandra} data accumulated over the past two decades. 
Specifically, close binaries often exhibit periodic modulations in their X-ray emission, which typically arise from binary orbital motion and/or spin of the accretor.  
Knowledge on the periodic signal, when complemented by spectral and multi-wavelength information, hold promise for revealing the nature of the X-ray source. 

In the case of CVs, the orbital period, which is controlled by the evolution of the donor in response to the mass transfer, is arguably the most powerful observational tool for studying CV formation and evolution. The standard model of CV evolution predicts the existence of a {\it period gap} between $\sim$ 2--3 hour and a {\it period minimum} at $\sim 80$ minute. 
These periodic features are related to the dominant angular momentum loss (AML) mechanism, which is responsible for driving the CV evolution and shrinking their orbits.
The AML is magnetic braking
for orbital periods $P_{\rm orb} \gtrsim$ 3 hr, whereas in short-period CVs ($P_{\rm orb} \lesssim$ 2 hr), the AML is dictated by gravitational radiation. Systems within the period gap have their mass transfer highly suppressed and thus few are observable. These periodic features are well supported by CVs found in the solar neighborhood \citep{2003A&A...404..301R}. 

Until recently, an orbital period has been found for only 15 CVs distributed in 6 GCs \citep{2012MmSAI..83..549K}. Nine of these were found in 47 Tuc, thanks to dedicated {\it Hubble Space Telescope} (HST) observations \citep[][hereafter E03a, E03b]{2003ApJ...596.1177E,2003ApJ...596.1197E}.
However, three of the eight periods were later rejected \citep[][hereafter RS18]{2018MNRAS.475.4841R}. 
This sample, albeit small and likely biased, exhibits an orbital period distribution in stark contrast with that of the field CVs, in the sense that most of these GC CVs have an orbital period above the period gap, while $\sim 70\%$ of the field CVs are found below the gap. 
This is considered tempting evidence that the GC CV population is substantially different from the field population, likely owing to dynamical interactions unique to the GC environment, as predicted by theoretical work \citep[e.g.,][]{2006ApJ...646..464S,2006MNRAS.372.1043I}. 
However, a firm conclusion on the role of dynamical effects must await for a statistically representative sample of GC CVs.

This motivates us to build a larger and less biased sample of GC CVs, and more generally of X-ray-emitting close binaries, which have a well-determined orbital period. 
In this work, we focus on 
47 Tucanae (= NGC 104), which is one of the most massive and best studied GCs in the Galaxy. Utilizing deep {\it Chandra} observations with an unparalleled resolving power, as well as the recently available eROSITA observations that cover the cluster outskirt, we conduct for the first time a systematic search for X-ray periodic sources in 47 Tuc. 
It is noteworthy that the resultant sample, like other flux-limited samples, may suffer from selection bias against short-period CVs, which tend to be fainter than their long-period counterparts in both the X-ray and optical bands.

This paper is organized as follows. In Section \ref{sec:data}, we describe the {\it Chandra} and eROSITA observations and our data reduction procedure. In Section \ref{sec:process}, we outline the basic principle of the period searching algorithm and the detailed applications to the {\it Chandra} and eROSITA data. 
The period searching results are presented in Section \ref{sec:clarification}, followed by a tentative classification of the periodic X-ray sources based on their temporal, spectral and multi-wavelength properties in Section \ref{sec:class}. 
Section \ref{sec:discussion} presents the orbital period distribution of the classified CVs, compares it with that of the field CVs, and addresses important implications on the dynamical formation of GC CVs. A summary of our study is given in Section \ref{sec:summary}.

\section{Data Preparation}
\label{sec:data}
\subsection{{\it Chandra} Observations}

The core region of 47 Tuc has been the target of {\it Chandra} with its Advanced CCD Imaging Spectrometer (ACIS) for 19 times. 
The first five observations, taken in 2000, were performed with the ACIS-I3 CCD on-axis.
The following 14 observations, taken between 2002 and 2015, were performed with the ACIS-S3 CCD on-axis, 
among which 13 were carried out with the sub-array mode to minimize photon pile-up from the bright sources. 
Table~\ref{tab:obsinfo} summaries the basic information of the 19 {\it Chandra}/ACIS observations
(see also Table 1 in \citealp{2019ApJ...876...59C} for more details).
These observations were utilized by \citet{2019ApJ...876...59C} to perform the most sensitive detection of X-ray sources in 47 Tuc to date, resulting in a catalog of 537 independent point sources, from which we search for periodic X-ray signals.
While these sources are found at a maximum projected distance of 7\farcm5 from the cluster center, the majority of them are located within the cluster's half-light radius of 3\farcm17, where the combined ACIS data have the highest sensitivity (see Figure 1 of \citealp{2019ApJ...876...59C}). 

We downloaded and uniformly reprocessed the archival data, using CIAO v.4.13 and calibration data files v.4.9.4 and following the standard procedure\footnote{http://cxc.harvard.edu/ciao}, to obtain the level-2 event file for each observation. Astrometry alignments among individual observations were fulfilled by matching the centroids of commonly detected points sources, assisted with the CIAO tool \emph{reproject\_aspect}. We used the observation with the longest exposure (69 ks), ObsID 2738, as the reference frame.
We further corrected the photon arrival time of each registered event to the Solar System barycenter (i.e., Temps Dynamique Barycentrique time) by using the CIAO tool \emph{axbary}.
We have examined the light curve of each observation and confirmed that the instrumental background was quiescent for the vast majority of time intervals. Thus, all the science exposures are preserved for the subsequent timing analysis, which ensures an uninterrupted light curve within each observation. 

Point-spread function (PSF) maps corresponding to a predefined enclosed count radius (ECR), later used for extraction of source light curve and spectrum, were generated for each observation, over the photon energy range of 0.5--8 keV. 
All PSF maps were weighted by a fiducial spectrum, which is an absorbed bremsstrahlung with a plasma temperature of 10 keV and a column density of $N_{\rm H}=2.3\times 10^{20}{\rm~cm^{-2}}$, representative of the X-ray sources in the cluster. 
We note that a
small fraction of the 537 sources catalogued by \citet{2019ApJ...876...59C} could be more significant
in a sub-band (e.g., 0.5--2 keV) than in the full 0.5--8 keV band. We
have examined sub-band light curves but found no extra significant periodic signals. Therefore we will focus on the 0.5--8 keV band for the {\it Chandra}-detected sources in the following.

\subsection{eROSITA Observations}
As one of the calibration targets, 47 Tuc was observed by eROSITA with the pointing mode for eight times shortly after its launch.
The basic information of these observations are listed in Table~\ref{tab:obsinfo}.
\citet{2021arXiv210614535S} carried out source detection over the combined field-of-view of the first five observations, taken on 2019-09-28, 11-01 and 11-02 and having individual exposures of 20--26 ks.
Three additional observations were taken on 2019-11-19, each with a shorter exposure of 8--9 ks. 

\begin{figure*}
 \centering
\includegraphics[scale=0.213]{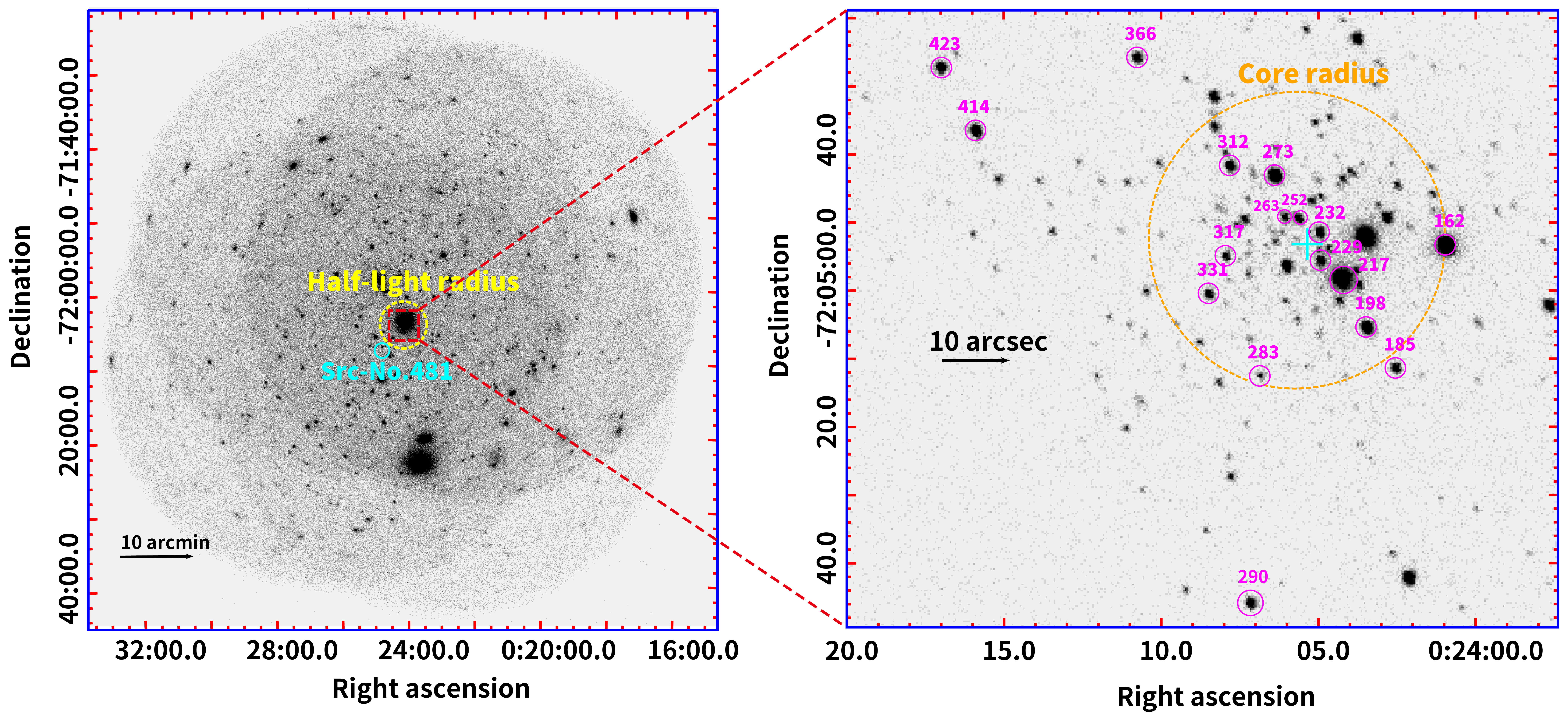}

\caption{{\it Left}: 0.2--5 keV counts image of the 47 Tuc field, combining 7 eROSITA observations. The dashed red box outlines a $2^\prime \times 2^\prime$ region containing the cluster core, which is unresolved due to source crowding and the limited angular resolution of eROSITA. The yellow circle represents the half-light radius of 3\farcm17. 
The only eROSITA source with a significant periodic signal is marked by a cyan circle and labelled following the source ID of \citet{2021arXiv210614535S}. {\it Right}: 0.5--8 keV {\it Chandra}/ACIS image of the $2^\prime \times 2^\prime$ region outlined in the left panel. The orange circle represents the core radius of 0\farcm36, and the cluster center is marked by the `+' sign. Seventeen {\it Chandra} sources with a significant periodic signal are marked by magenta circles and labelled following the IDs of \citet{2019ApJ...876...59C}.}
\label{fig:fov}
\end{figure*}

We downloaded and reprocessed the data using the eROSITA Science Analysis Software System (eSASS; \citealp{2021A&A...647A...1P}).
We applied the barycentric correction for photon arrival times, based on the publicly released orbit file for individual observations\footnote{https://erosita.mpe.mpg.de/edr/eROSITAObservations/OrbitFiles} and utilizing the {\it barycen} task that is part of the HEASoft (v6.29) package.
Unfortunately, orbit file is not available for the very first observation, obsID=700012, which may introduce substantial uncertainty in the photon arrival time. Thus we discarded this observation and used the remaining seven observations for subsequent timing analysis.
The left panel of Figure~\ref{fig:fov} displays the combined field-of-view of the eROSITA observations.

Similarly, we generated 0.2--5 keV PSF maps for each observation using the eSASS task \emph{apetool}, according to which the source light curve and spectrum were then extracted. We adopt the source list of \citet{2021arXiv210614535S}, which includes 888 sources detected over 0.2--5 keV and located within a projected distance of $42^\prime$ from the cluster center, whose centroid positions have been registered to the more accurate {\it Chandra} positions provided by \citet{2019ApJ...876...59C}.
Due to the moderate angular resolution of eROSITA and source crowding in the cluster core, 
\citet{2021arXiv210614535S} excluded the central $1\farcm7$ region, which is fortunately well resolved by the {\it Chandra} observations (right panel of Figure~\ref{fig:fov}).
We note that \citet{2021arXiv210614535S} classified 92 background active galactic nuclui (AGNs) and 26 Galactic foreground stars, chiefly on the basis of tentative optical counterparts. These sources are included in our timing analysis for completeness. 

\begin{table}
\centering
\caption{{\it Chandra} and eROSITA observations of 47 Tuc} 
\label{tab:obsinfo}
\centering
\begin{tabular}{lcccr}
\hline
\hline
ObsID & Date & R.A. & Decl. & Live Time  \\
(1) & (2) & (3) & (4) & (5)   \\ 
\hline
{\it Chandra} &  & &  &\\
\hline
78    & 2000-03-16  & 5.97704 & -72.07297 & 3.87  \\
953   & 2000-03-16  & 5.97695 & -72.07304 & 31.67 \\
954   & 2000-03-16  & 5.97716 & -72.07304 & 0.85 \\
955   & 2000-03-16  & 5.97696 & -72.07294 & 31.67 \\
956   & 2000-03-17  & 5.97700 & -72.07282 & 4.69  \\
2735  & 2002-09-29  & 6.07523 & -72.08251 & 65.24  \\
2736  & 2002-09-30  & 6.07516 & -72.08274 & 65.24  \\ 
3384  & 2002-09-30  & 6.07515 & -72.08263 & 5.31  \\
3385  & 2002-10-01  & 6.07498 & -72.08296 & 5.31  \\
2737  & 2002-10-02  & 6.07491 & -72.08330 & 65.24  \\
3386  & 2002-10-03  & 6.07480 & -72.08349 & 5.54  \\
2738  & 2002-10-11  & 6.07322 & -72.08552 & 68.77  \\
3387  & 2002-10-11  & 6.07299 & -72.08544 & 5.73   \\
16527 & 2014-09-05  & 6.01654 & -72.07804 & 40.88  \\
15747 & 2014-09-09  & 6.01653 & -72.07805 & 50.04   \\
16529 & 2014-09-21  & 6.01989 & -72.07845 & 24.7  \\
17420 & 2014-09-30  & 6.01989 & -72.07840 & 9.13   \\
15748 & 2014-10-02  & 6.01984 & -72.07846 & 16.24  \\
16528 & 2015-02-02  & 6.01851 & -72.08395 & 40.28  \\
\hline
{\it eROSITA} & & & &\\
\hline
700012$^\dag$ & 2019-09-28 & 6.5339 & -72.1763 & 19.5 \\
700011 & 2019-11-01 & 5.5180 & -71.9862 & 25.8 \\
700163 & 2019-11-02 & 6.5339 & -72.1763 & 25.3 \\
700013 & 2019-11-02 & 6.3338 & -71.9249 & 25.2 \\
700014 & 2019-11-02 & 5.7108 & -72.2375 & 25.2 \\
700173 & 2019-11-19 & 5.5180 & -71.9862 & 8.8  \\
700174 & 2019-11-19 & 6.5339 & -72.1763 & 8.4  \\
700175 & 2019-11-19 & 6.3338 & -71.9249 & 8.3  \\
\hline
\hline
\end{tabular}
\begin{tablenotes}
      \small
      \item 
      Notes:
      (1)-(2) Observation ID and date.
      (3)-(4) J2000 sky coordinates of the telescope aimpoint, in degrees. (5): Effective exposure time, in units of ks.
	  $^\dag$This observation is not used due to the lack of satellite orbital information. 
\end{tablenotes}
\end{table}

\section{Period Searching Procedure}
\label{sec:process}
\subsection{The Gregory-Loredo algorithm} \label{subsec:GL}

The Gregory-Loredo (GL) algorithm \citep{1992ApJ...398..146G}, which belongs to the category of phase-folding methods, applies the Bayesian probability theory to evaluate the likelihood of a periodic model against a model of constant flux and subsequently determines the most probable period, $P$.  
The key of this algorithm is the multiplicity of the phase distribution of events,
\begin{equation}\label{multi}
W_m(\omega, \phi)={{N!}\over{n_1!\; n_2!\; n_3!\cdots {n_m}!}},
\end{equation}
where $N$ represents the total number of counts, 
$n_i(\omega, \phi)$ is the number of counts falling into the $i$th of $m$ phase bins, given the frequency $\omega = 2\pi/P$ and phase $\phi$, satisfying $\sum\limits_{i=1}^{m}n_i(\omega, \phi)=N$. 
The multiplicity is the number of ways that the binned distribution could have arisen by chance and the more $n_i$ differ from each other, the smaller the multiplicity. In other words, the more the stepwise model defined by the $m$ phase bins deviates from constant, the more likely there exists a periodic signal, the probability of which is inversely proportional to the multiplicity. 
A brief summary in mathematical form of the Bayes's theorem and the GL algorithm can be found in Appendix A of \cite{2020MNRAS.498.3513B}.

The GL algorithm has recently been applied to 1-Ms {\it Chandra} observations of a Galactic bulge field (the so-called Limiting Window; 
\citealp{2020MNRAS.498.3513B}) and the 7-Ms {\it Chandra} deep field south \citep{2022MNRAS.509.3504B} to detect periodic X-ray sources. 
These studies demonstrated the reliability and efficiency of the GL algorithm for probing periodic signals in X-ray sources with a moderate number of counts and sampled by an irregular observing cadence, which is typical of X-ray observations.
Here we follow the recipe outlined in \citet{2020MNRAS.498.3513B} to perform a systematic search for periodic signals from the X-ray sources in the {\it Chandra} and eROSITA fields of 47 Tuc. 
We adopt a probability of 90\% as the threshold for tentative periods returned by the GL algorithm.

\subsection{Application to the Chandra \& eROSITA data} 
\label{subsec:appli}

For each {\it Chandra} source, we extract the 0.5--8 keV counts within the 90\% ECR of individual observations to form a time series as input to the GL algorithm.
In the crowding cluster core, the default 90\% ECR of some neighboring sources overlap; for such cases we adopt the 75\% ECR for source count extraction. 
Since the GL algorithm manipulates each photon arrival time to evaluate the probability of periodic variation against a constant model, there is no need to separately account for the background level, which is absorbed into the presumed constant. 
Nevertheless, we estimate the background level for each source by extracting counts from within a concentric annulus with inner-to-outer radii of 2--4 times the 90\% ECR, masking any pixel falling within two times the 90\% ECR of neighbouring sources.

By design, the GL algorithm folds the time series at trial frequencies (periods). The resolution and range of searched period is compromised between efficiency and computational power. 
Following \citet{2020MNRAS.498.3513B}, we restrict our search in three period ranges: (100, 3000), (3000, 10000) and (10000, 50000) sec, with a frequency resolution of $10^{-7}$, $10^{-8}$ and $10^{-9}$ Hz, respectively. 
Given the timespan of $\sim 5 \times 10^8$ sec between the first and last {\it Chandra} observations (Table~\ref{tab:obsinfo}), the chosen frequency resolutions are optimal for an efficient search of periodic signals.
Since the GL algorithm only determines the most probable period, 
once a tentative period is identified, a second search is performed excluding a narrow interval around the identified period, which ensures that a possible second period within the same period searching range will not be missed. 

The chosen period ranges are optimal for detecting the orbital and spin periods of CVs. The orbital period distribution of CVs is known to exhibit a minimum at $\sim$82 min and a gap between $\sim$2--3 hours \citep{2011ApJS..194...28K}. CVs typically have an orbital period shorter than 10 hours in order to meet the Roche lobe filling condition, but in rares cases which involve an evolved donor, the orbital period can be significantly longer \citep{2016ApJ...833...83K}. The second and third period ranges cover these characteristic periods, whereas the first range probes the spin period of fast rotating WDs in IPs.
We note that ABs and LMXBs, as well as CVs in {\bf rare} cases, can have an orbital period longer than 50 ks. However, the current X-ray data of 47 Tuc, with individual exposures $\lesssim70$ ks (Table~\ref{tab:obsinfo}), are not optimal for detecting such long periodic signals without facing false alarms due to long-term aperodic variability often present in these X-ray sources (see more discussions below).

As for the eROSITA sources, time series are extracted from within the 75\% ECR, which is sufficiently large to include most source counts and also sufficiently small to avoid source overlapping, given the large PSF of eROSITA compared to that of {\it Chandra}. 
It turns out, however, most eROSITA sources have a significantly varied 75\% ECR from observation to observation, due to their varied off-axis positions in the seven observations, which have largely separated aimpoint positions (Figure~\ref{fig:fov} and Table~\ref{tab:obsinfo}).
The increased ECR at large off-axis angles necessarily captures more background counts into the source aperture, resulting in an artificially time-variable light curve (see Figure~\ref{fig:erosita_src} for an example). 
We note that for most of the {\it Chandra} sources the background level is too low to cause a variable light curve.
As noted by \citet{2020MNRAS.498.3513B}, a light curve with substantial aperiodic variability fed to the GL algorithm may result in a fake signal, typically at a period comparable to the timescale of the variation, if too many photons happened to fall within a small fraction of phase bins.
Thus we only analyze time series from the seven individual observation.
For the four $\sim$25 ks-long observations, we choose (100, 3000) and (3000, 10000) sec as the period searching range, with a frequency resolution of $10^{-7}$ Hz,
while for the three $\sim$8 ks-long observations, only the range of (100, 3000) sec is searched for periodic signals.
We also apply the GL algorithm to the combined time series from all seven observations to search for signals in the range of (10000, 50000) sec.

To provide a reliability check of any long-period signal reported by the GL algorithm on the eROSITA data,
we further apply the generalized Lomb-Scargle (hereafter LS) periodogram \citep{1976Ap&SS..39..447L,1982ApJ...263..835S}, with a normalization of sample variance following \citet{2009A&A...496..577Z}. 
The LS periodogram probes periodic signal from a background-subtracted light curve, hence it suffers little from the aforementioned effect of strongly varied PSF.
The LS periodogram has its own drawback, in the sense that it does not perform well with interrupted light curves having wide gaps in between.
Nevertheless, the four 25-ks eROSITA observations were taken nearly continuously.
Therefore, we apply the LS periodogram to background-subtracted light curves extracted from these four observations, which have a combined length of $\sim100$ ks. 
The background count rate in each of the four observations is estimated from the smoothed background image generated by the eSASS tool \emph{erbackmap}\footnote{https://erosita.mpe.mpg.de/edr/DataAnalysis/erbackmap\_doc.html}. 

We apply the false alarm probability proposed by \citet{2008MNRAS.385.1279B} to assess the significance of the peak in  the LS periodogram (i.e., a tentative period), the robustness of which was demonstrated by \citet{2022MNRAS.509.3504B}. A valid detection is defined as having a false alarm probability lower than 0.27\%, i.e., a $3\sigma$ significance.

Given the expectation that 47 Tuc may host abundant NS-LMXBs, we have also extended our period searching to the range of (10, 100) sec, with a resolution of $10^{-6}$ Hz, to probe pulsating signal from an NS, 
However, no significant signal in this period range is found in either {\it Chandra} or eROSITA sources.

\section{Period searching results}
\label{sec:clarification}
\subsection{Result of {\it Chandra} sources}
\label{subsec:reschandra}

From the time series of the 537 {\it Chandra} sources, the GL algorithm reports 41 candidate periodic signals having a probability greater than 90\%. 
However, some of these signals may be spurious, due to one of the following reasons:

(i) The {\it Chandra}/ACIS operates in a dithering pattern, with a period of 706.96 s in pitch and 999.96 s in yaw{\footnote{https://cxc.harvard.edu/ciao4.4/why/dither.html}}, to distribute the photons over more CCD pixels. Any signals detected at these two periods or their harmonics to within 1\% are considered spurious. We thus exclude 9 such spurious signals, all found in sources located near CCD gaps or edges. 

(ii) As mentioned before, an aperiodically variable light curve may fool the GL algorithm to report a false period. 
For each source with a candidate period, we inspect its light curve to identify short-term (i.e., intra-observation) flares.
The situation happens in four sources, producing 6 fake periodic signals mainly around the flare duration or the length of the affected observation.
Moreover, for sources exhibiting strong long-term (i.e., inter-observation) variations, we repeat the period search in two subsets of the light curve: one covering only the high state (defined as the observation[s] with the highest photon flux) and the other excluding the high state.   
If the periodic signal could not be reproduced in either subset, we consider it a false detection.
In this way we exclude 7 tentative periods, all above 10 hours. 

Moreover, we note a well-known ambiguity in distinguishing the true period and its harmonics and sub-harmonics, i.e., integer division or multiplication of the true period{\footnote{It is noteworthy that harmonics are conventionally defined in frequency space. For convenience and not losing clarity, throughout this work we use the term harmonics on the period space.}}.
In principle, the light curve can tell the true period only when we fully understand the mechanism behind the periodic variability. For example, polars often exhibit single-peaked structure in their light curves due to the accretion dominated by one of the two magnetic poles. In this case, the harmonics can be recognized by a double-peaked shape from the phase-folded light curve.
On the other hand, a double-peaked shape modulated by the true period may exist in the light curve of IPs, in which case the accretion-induced X-ray emission can arise from both magnetic poles \citep{2020MNRAS.498.3513B}.
With this caveat in mind, we have inspected the likelihood of the detected periods being (sub-)harmonics of the true period on a best-effort basis, assisted with literature information typically derive from optical and ultraviolet observations (see details in Section~\ref{subsec:classification}). 
This has led us to recognize that one periodic signal, found in Seq.229 (source sequence number following \citealp{2019ApJ...876...59C}), is in fact an integer division of the true period at 95371.20 s, which was previously unambiguously identified from orbital eclipse in this source \citep{2003ApJ...599.1320K}. Due to the aforementioned limitation in the period searching ranges, the GL algorithm originally reports a period at about one-third of the true value. Once we expand the searching range, the GL algorithm recovers the true period, which is clearly due to orbital eclipse (Figure~\ref{fig:pfold_lc}).
We find no strong evidence for a confusing (sub)-harmonic in the other sources.

After the above filtering, there remain 19 valid periodic signals found among 17 {\it Chandra} sources, among which two sources each exhibit dual periods. The basic information of these sources are summarized in Table \ref{tab:srcinfo}, in the order of increasing periods. All but two periods have a GL probability greater than 99\% (all above 96\%).  
The positions of these sources are indicated in the right panel of Figure~\ref{fig:fov}.
Classification of these sources will be addressed in Section~\ref{sec:class}.

\subsection{Result of eROSITA sources}\label{subsec:reseROSITA}

As stated in Section \ref{subsec:appli}, the application of the GL algorithm to the eROSITA data is restricted to either single observations for periodic signals shorter than 10 ks, or the time series combining all seven observations for periodic signals between 10--50 ks.
The latter case also requires confirmation with the LS periodogram.

\begin{figure*}
\centering
\includegraphics[scale=0.5]{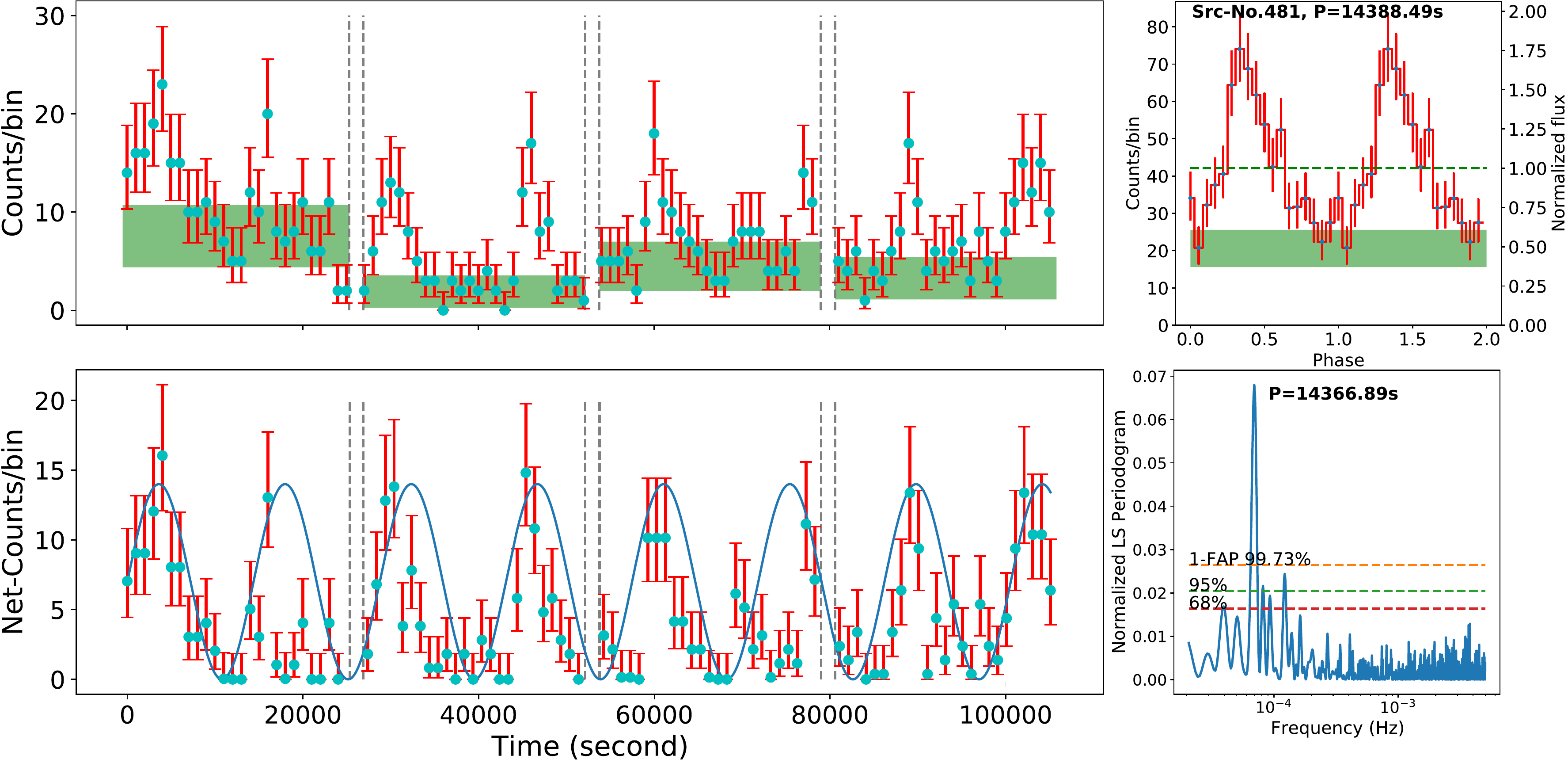}
\caption{
{\it Upper left:} The eROSITA 0.2--5 keV light curve of Src-No.481 binned by 1000 seconds. The grey vertical dashed lines label the gap between observations, whereas the green strip represents the local background, the width of which represents 1\,$\sigma$ Poisson error.
{\it Upper right:} Phase-folded light curve at the period determined by the GL algorithm. The green dashed line represents the mean count rate.
{\it Lower left:} The eROSITA background-subtracted 0.2--5 keV light curve, overlaid by a sinusoidal curve at the most probable period determined by the LS periodogram.
{\it Lower right:} Normalized LS periodogram of Src-No.481. Three false alarm levels are denoted by the colored dotted lines.}
\label{fig:erosita_src}
\end{figure*}

In the first case, 
after filtering spurious signals due to strong aperiodic variability,
only one source, Src-No.481 in \citet{2021arXiv210614535S}, is identified by the GL algorithm, which exhibits a periodic signal at a period of $\sim$ 6792 s and $\sim$ 7825 s in the observation of ObsID 700163 and ObsID 700013, respectively. 
A periodic signal around 7 ks is also evident in the other two 25-ks observations, but with a lower significance. 
In the latter case, only one source, again Src-No.481, is identified by both GL and LS to exhibit a period of 14388.49 s and 14366.89 s, respectively.
The upper panel of Figure \ref{fig:erosita_src} displays the raw light curve fed to the GL algorithm, along with the phased-folded light curve according to $P = 14388.49$ s,
and the lower panel displays the background-subtracted light curve fed to the LS periodogram, along with a sinusoidal curve with $P = 14366.89$ s, which provides a reasonable characterization to the observed light curve, in particular being able to match all but one of the alternating peaks. 
This suggests that the 7--8 ks periodic signal found in the individual observations is more likely due to a harmonic at half of the true  period, which is taken to be 14388.49 s.
It is noteworthy that although this source is also detected by {\it Chandra} (Seq.503 in \citealp{2019ApJ...876...59C}), the number of ACIS-detected counts is insufficient to recover the periodic signal found in the eROSITA data.  

As illustrated by Figure \ref{fig:fov}, all 17 {\it Chandra} sources with a periodic signal are located within the central $1\farcm7$ region of 47 Tuc, where eROSITA can barely resolve  individual sources due to source crowding. Nevertheless, for the {\it Chandra} periodic sources, it is still possible to confirm the periodic signal with the eROSITA data, provided that background noise does not lead to a complete dilution. 
Hence we extract eROSITA time series  from the positions of the 17 {\it Chandra} sources within a 50\% ECR and run the GL algorithm. 
Only one source, Seq.162, with a $\sim$ 31200s signal, 
is recovered by the eROSITA data. 
Figure~\ref{fig:src_example} shows the phase-folded light curves from both telescopes, in which the substantially higher background level of eROSITA is obvious.
The periodic signal is clearly due to an eclipse, during which the {\it Chandra}-detected flux completely vanishes. 
The eROSITA light curve, after accounting for its different chronological zero point\footnote{https://erosita.mpe.mpg.de/edr/DataAnalysis/prod\_descript/header\\-keywords\_edr.html} with respect to that of {\it Chandra}, finds perfect phase-matching with the {\it Chandra} light curve, despite a separation of over 4720 phase cycles between the two. 
This consistency lends strong support to the validity of our period searching process with the eROSITA data.

\begin{figure*}
\centering
\includegraphics[scale=0.9]{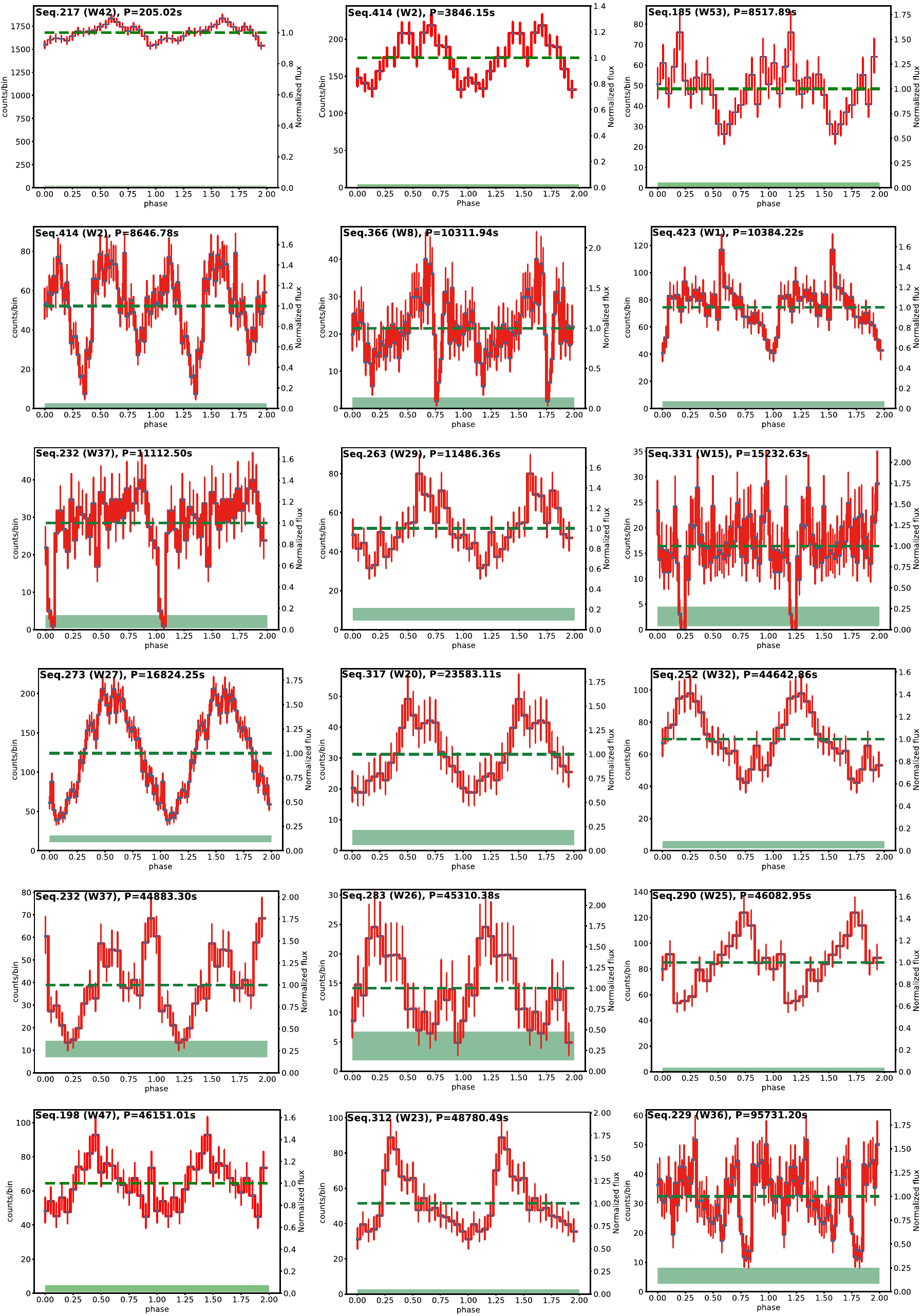}

\caption{The 0.5--8 keV phase-folded light curve at the modulation period.
The green dashed line represents the mean count rate, whereas the green strip represents the local background, the width of which represents 1\,$\sigma$ Poisson error.}
\label{fig:pfold_lc}
\end{figure*}

\begin{figure}
\includegraphics[width=0.49\textwidth]{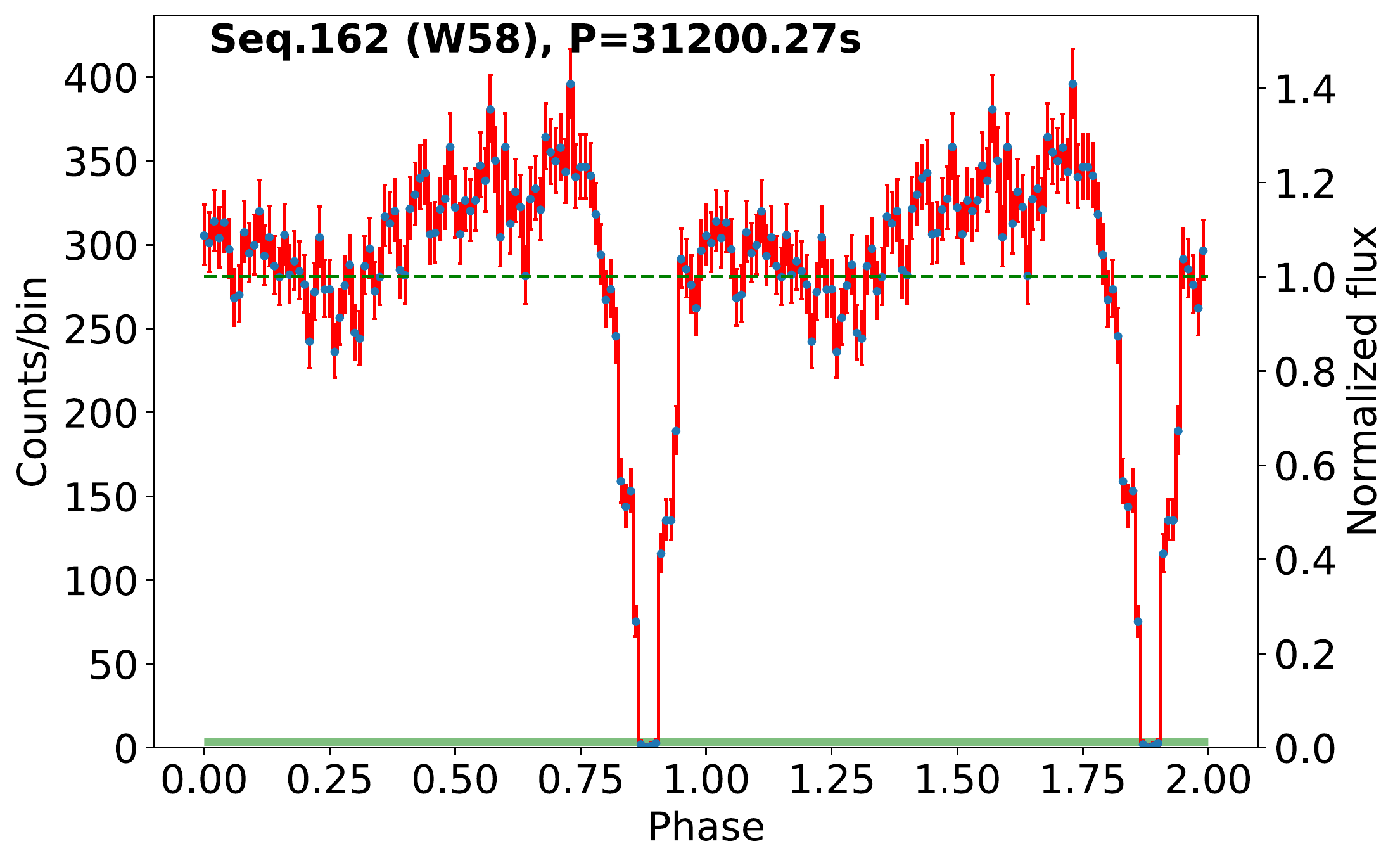}
\includegraphics[width=0.49\textwidth]{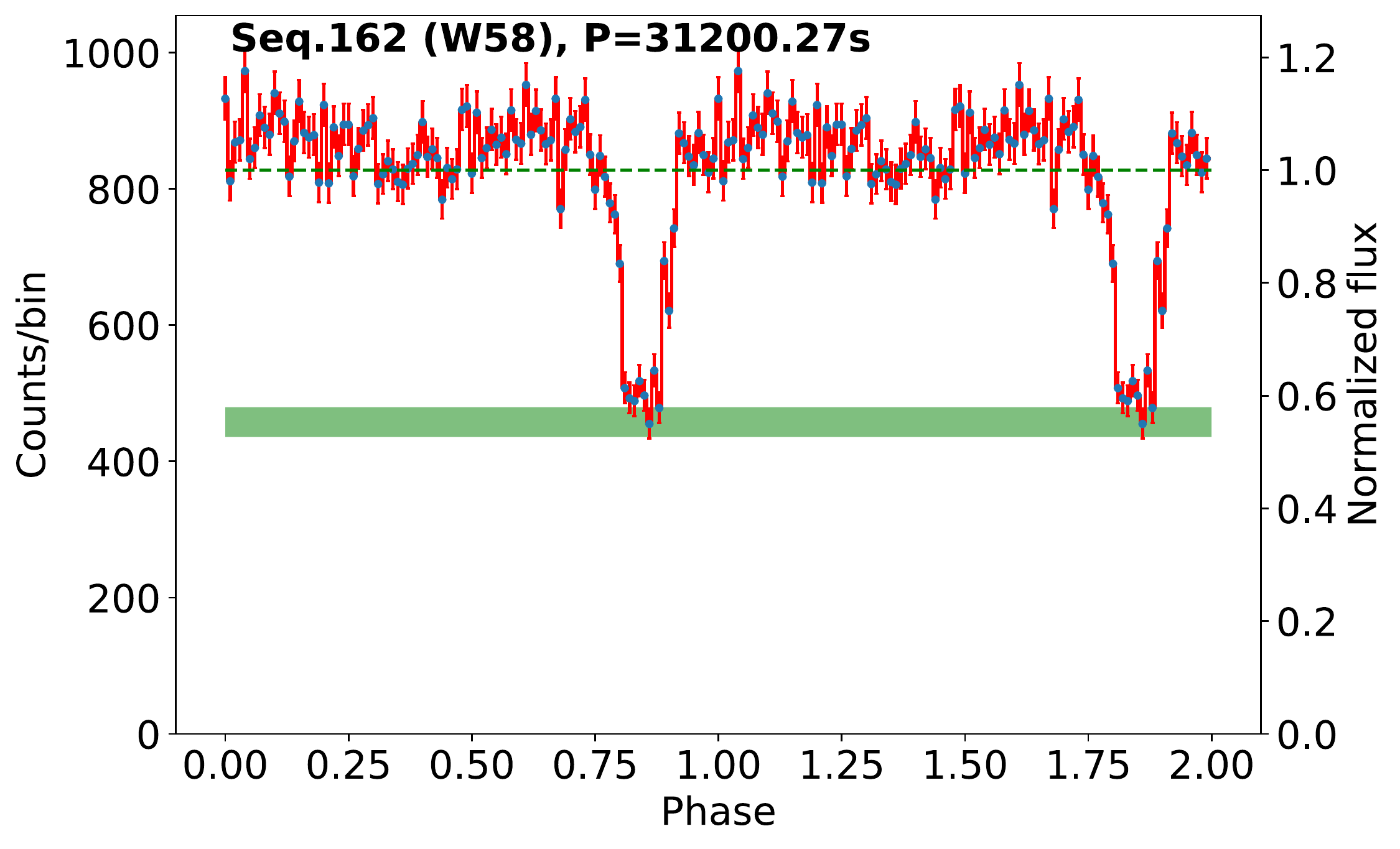}
\caption{{\it Upper panel:} The phase-folded light curve of source Seq.162, from the {\it Chandra} observations. 
{\it Lower panel:} The phase-folded light curve at the same period, from the eROSITA observations. 
The green dashed line represents the mean count rate, whereas the green strip denotes the local background, the width of which represents 1-$\sigma$ Poisson error. The phase of both light curves has been registered to the common time-zero point of MJD = 51543.875.}
\label{fig:src_example}
\end{figure}

To summarize, we have identified 18 X-ray sources with a total of 20 periodic signals, based on the data of the two telescopes. 
Phase-folded light curves of these sources, except for Src-No.481 (Figure \ref{fig:erosita_src}) and Seq.162 (Figure \ref{fig:src_example}), are provided in Figure~\ref{fig:pfold_lc}.

\begin{table*}
\tabcolsep=0.165cm
\centering
\caption{Basic information of the periodic X-ray sources in 47 Tuc}\label{tab:srcinfo}
\begin{threeparttable}
\begin{spacing}{1.5}
\begin{tabular}{lllllllcllclll}
\hline
\hline
Seq    & RA       & DEC   & $R$   & $P$ & $\sigma_{\rm P}$ & Prob.   & $P_{\rm opt}$   & $C$ & $C_{\rm B}$ & FAP & Eclipse & Class & Note    
\\ 
  & deg & deg & arcsec & second & \% &  & second & counts & counts & \% &   
\\
(1) & (2) & (3) & (4) & (5) & (6) & (7) & (8) & (9) & (10) & (11) & (12) & (13) & (14)
\\
\hline

217 & 6.01780 & -72.08281  & 8.5 & 205.02 & 0.02 & 0.9999 & -  & 33611  & 144.7 & 0.0 & No & LMXB & W42, X9, UV            \\

$414^{\dag}$  & 6.06625 & -72.07680  & 49.9 & {\bf 3846.15} & 0.02 & 0.9982 & - & 2620 & 20.8 & 0.0
  & No & CV  & W2, X13, UV    \\

185 & 6.01063 & -72.08649  & 23.6 & {\bf 8517.89} & 0.10 & 0.9987 & - & 966 & 35.8 & 0.0  & Yes & CV &  W53, UV                        \\

$414^{\ddag}$ & 6.06625 & -72.07680  & 49.9 & 8646.78 & < 0.01  & 1.0000 & 8649.00 & 2620 & 20.8 & 14.3  & Yes & CV & W2, X13, UV        \\

366    & 6.04485  & -72.07383 & 35.7 & {\bf 10311.94} & 1.17 & 1.0000 & 10304.93 & 1075  & 16.1 & 1.7  & Yes & CV & W8, UV               \\

423    & 6.07077  & -72.07424  & 58.1 & {\bf 10384.22} & 0.35 & 0.9986  & 20779.20  & 2231   & 17.3 & 0.0  & Yes & CV & W1, UV            \\

$232^{\dag}$    & 6.02075  & -72.08095  & 3.4 & 11112.50 & 0.68 & 0.9999 & - & 1664  & 132.9 & 0.0  & Yes & LMXB & W37           \\

263    & 6.02533  & -72.08032  & 3.9 & 11486.36 & 0.66 & 0.9662 & 11486.02   & 569    & 70.4  & 0.0  & No & LMXB & W29, MSP         \\

e481 & 6.18225  & -72.13861 & 270.9 & {\bf 14366.89} & 1.96 & 1.0000 & -  & 784 & 372.6 & 43.4  & No & MSP &  Seq.503\\

331    & 6.03540 & -72.08345 & 15.2 & {\bf 15232.63} & 1.44 &  0.9849 & 15240.96 & 828 & 20.2 & 0.0  & Yes & CV & W15, UV      \\

273    & 6.02666  & -72.07863 & 10.1  & 16824.25 & 0.16 &  1.0000  & 13824.00  & 6218   & 30.2 & 95.6  & No & CV & W27, X10, UV  \\

317    & 6.03316  & -72.08192 & 10.8   & {\bf 23583.11} & 1.48 & 0.9945 & - & 629 & 22.6 & 0.0 & No & CV & W20   \\

162 & 6.00405  & -72.08147 & 21.7 & 31200.27 & 0.17 & 1.0000 & - & 28020  & 69.3 & 100.0 & Yes & LMXB & W58, X5          \\

252    & 6.02362  & -72.08036 & 3.3 & {\bf 44642.86} & 1.28 & 0.9999 & 22947.84   & 1001   & 152.5 & 0.6 & No & CV & W32 \\

$232^{\ddag}$   & 6.02075  & -72.08095  & 3.4 & {\bf 44883.30} & 1.27 & 1.0000 & - & 1664  & 132.9 & 65.2 & Yes & LMXB & W37           \\

283    & 6.02860  & -72.08681 & 20.7 & {\bf 45310.38} & 2.14 & 0.9999 & 34029.50  &  287  &  28.2 & 0.4 & No & AB & W26           \\

290    & 6.029851  & -72.09607 & 53.7 & {\bf 46082.95} & 1.88 & 0.9912  & - &  1272  &  13.7 & 1.2 & Yes & CV & W25, UV           \\

198 & 6.01452  & -72.08482  & 16.2 & {\bf 46151.01} & 1.56 & 1.0000 & 45835.20 & 3517   & 47.7 & 0.8 & No & AB & W47, E8 \\

312    & 6.03259  & -72.07823 & 14.8 & {\bf 48780.49} & 0.15 & 0.9999 & 22242.82 & 1618   & 22.0  & 10.4 & No & CV & W23, UV  \\        

229    & 6.02057  & -72.08210 & 4.5 & {\bf 95731.20} & 1.93 & 1.0000 & 95731.20 & 1638   & 244.1  & 3.9 & Yes & CV & W36, AKO-9, UV    \\

\hline
\end{tabular}
\end{spacing}
\begin{tablenotes}
      \small
      \item
      Notes: 
      (1) Source sequence number taken from \citet{2019ApJ...876...59C}, except for the eROSITA source e481, which is from \citet{2021arXiv210614535S}.
      The same source with dual periods is marked by \dag\ and \ddag.
(2) and (3) Right Ascension and Declination (J2000) of the source centroid. 
(4) The projected distance from the cluster centre. 
(5) The modulation period determined by the GL algorithm. Newly detected periods in the X-ray band are highlighted in bold face. 
(6) The estimated relative uncertainty of the period, expressed in percentage.
(7) The GL probability.
(8) Period found in the optical and reported by \citet{2003ApJ...596.1177E},  \citet{2001ApJ...559.1060A} or \citet{2018MNRAS.475.4841R}.
(9) The number of total counts in the 0.5--8 keV band.
(10) The number of estimated background counts.
(11) False alarm probability due to potential red noise, as described in Section~\ref{subsec:rednoise}.
(12) The presence/absence of eclipsing behavior.
(13) Tentative source classification.
(14) Notes for alias (mainly from \citealp{2001Sci...292.2290G}), possible UV counterpart \citep{2018MNRAS.475.4841R} and possible MSP counterpart. 
\end{tablenotes} 
\end{threeparttable}
\end{table*}

\subsection{Potential caveat due to red noise}
\label{subsec:rednoise}
Accretion-powered systems, such as CVs, LMXBs and active galactic nuclei (AGNs), are known to exhibit aperiodic variability on a wide range of time scales \citep{1950HarCi.455....1L,1971ApJ...168L..43R,1974PASJ...26..303O}. 
The so-called red noise, a main component of aperiodic variation, may cause false periodic signal especially at lower frequencies \citep{1989IBVS.3383....1W}. 
We estimate the probability of the GL algorithm being false-alarmed by aperiodic variation, using a similar procedure as in
 \citet{2022MNRAS.509.3504B}.
Specifically, we first adopt an analytical model of the source power spectrum,
which has the form of,
\begin{equation}
P(\nu) = N \nu^{-1}\left(1+\left(\frac{\nu}{\nu_0}\right)^4\right)^{-1 / 4}+C_{\rm P},
\label{eqn:CVps}
\end{equation}
or
\begin{equation}
P(\nu)=N \nu^{-\alpha}+C_{\rm p}.
\label{eqn:LMXBps}
\end{equation}
Eq.~\ref{eqn:CVps} was proposed by \citet{2010A&A...513A..63R} for describing the power spectra of CVs \citep{2012A&A...546A.112B},
while Eq.~\ref{eqn:LMXBps}
is suitable for the power spectra of LMXBs.
Here $N$ is the normalization factor, $\alpha$ is the spectral index, and $C_{\rm P}$ represents the Poisson noise determined by the mean photon flux of the source.
The break frequency $\rm \nu_0$ is usually $\gtrsim$ 1 mHz for CVs and cannot be well constrained by the current data due to the presence of Poisson noise. Thus we fix $\rm \nu_0$ at 1 mHz, which has little effect in the results.
To mitigate the potential effect of interrupted observations in the Fourier analysis, we take the longest  {\it Chandra} observation, i.e., ObsID 2738, to characterize the power spectrum for each of the 18 periodic sources. 
We have also fitted the power spectrum using the other long observations (ObsIDs 2735, 2736, 2737), but found no significant difference for most sources. The only exceptions are Seq.162 and Seq.273, which show a varying power spectrum among the different observations. Nevertheless, for uniformity we stick with ObsID 2738 to constrain the power spectrum of these two sources, which has little effect in the following estimation of the false alarm probability.
We have also masked a narrow frequency window around the detected period of a given source. 
As for Src-No.481, we characterize its power spectrum using all four long eROSITA observations, because the observation gap is negligibly small.

\begin{figure}
\centering
\includegraphics[scale=0.45]{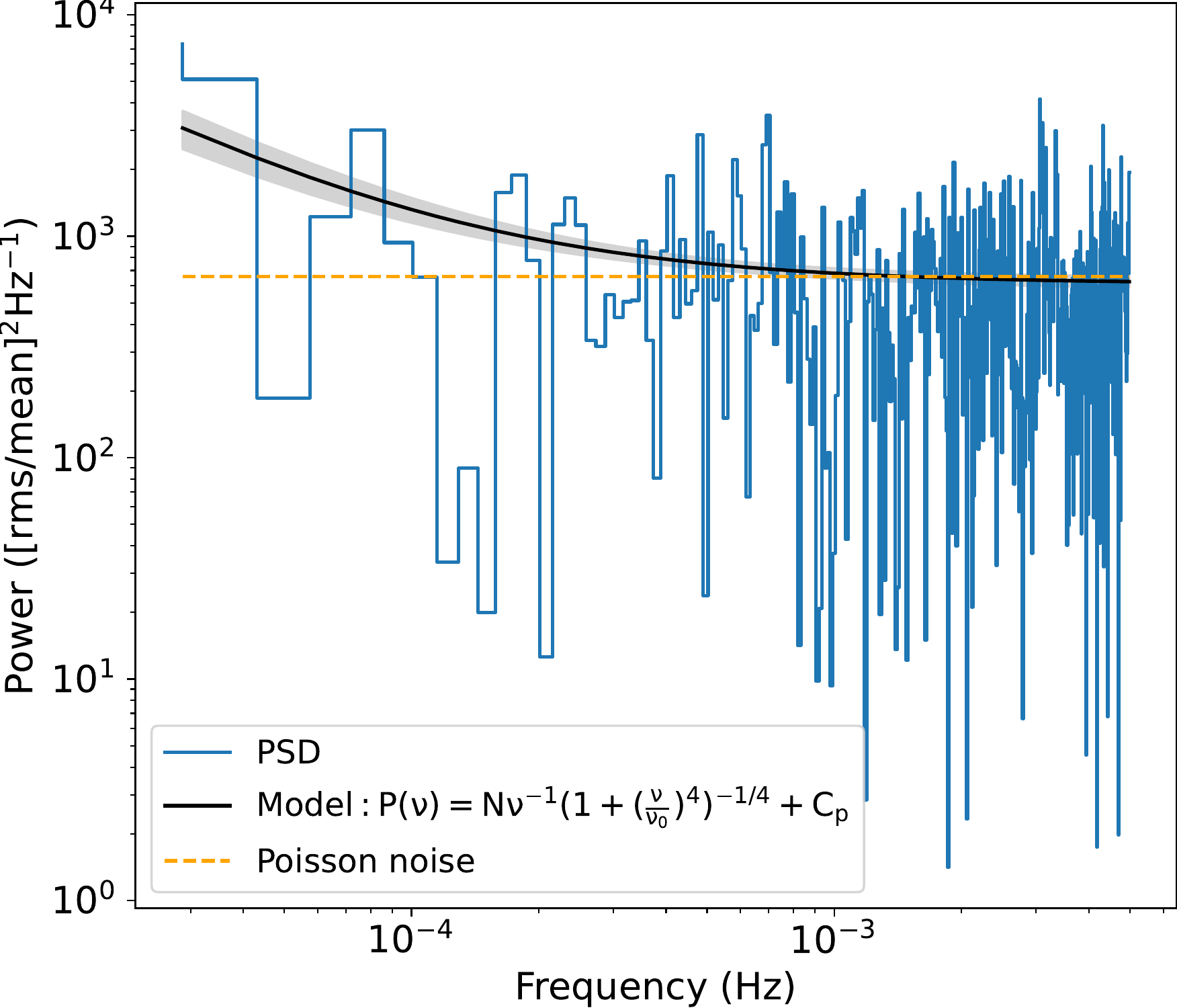}

\caption{The power spectral distribution (blue histogram), normalized to $\rm (rms/mean)^2$ per Hz, of source Seq.312 in ObsID 2738.
The best-fit model and 1$\sigma$ error are presented by the black solid line and the grey strip, while the theoretical Poisson noise level is marked by the orange dashed line.} 
\label{fig:312psd}
\end{figure}

The power spectrum of a given source is fitted (with Eq.~\ref{eqn:CVps} or Eq.~\ref{eqn:LMXBps} dependent on the preferred source class; see Section~\ref{sec:class}) using the maximum likelihood function discussed in \cite{2010MNRAS.402..307V} and the Markov Chain Monte Carlo approach (with the python \emph{emcee} package, \citealp{2013PASP..125..306F}) to determine the best-fit parameters and errors. 
Then we simulate the time series based on the best-fit model, following the method proposed by \citet{1995A&A...300..707T}. 
An example of the fitted power spectrum is shown in Figure~\ref{fig:312psd}.

For each source, a photon time series is simulated for the same epochs of the actual observations, which are then fed to the GL algorithm, with the same period searching range as applied in Section \ref{subsec:appli}. A group of 1000 simulated time series are produced for each source to evaluate the false alarm probability (FAP) resulted from red noise, i.e., the fraction of these 1000 time series leading to a false detection. 
We define a false detection as a signal reported with a GL probability above 0.99, which is consistent with the majority of the deteted periodic signals (Table~\ref{tab:srcinfo}). 
To be conservative, we refer to a ``global'' FAP, that is, all reported signals falling in a given period searching range are taken as false detections. Except for Seq.414 and Seq.232, both are detected with dual periods in the same range, we  divide the given period searching range to two parts, to product FAPs for each periods.

The resultant FAP for each detected periodic signal is reported in Table~\ref{tab:srcinfo}. 
A low FAP ($\lesssim4\%$) is found for most periodic signals, indicating that they are unlikely false detections due to red noise. Six signals (Src.-No.481, Seq.273, Seq.162, Seq.312, the longer period of Seq.414, and the longer period of Seq.232), on the other hand, have an FAP above 10\%. 
However, it should be emphasized that a high FAP is {\it not} a sufficient condition for a fake periodic signal. 
For instance, Seq.162 has the highest FAP (100\%) due to its strong red noise, which results in fake signals at low frequencies (long periods), but its detected period is clearly a {\it true} period arising from orbital eclipse (Figure~\ref{fig:src_example}).
Similarly, the phase-folded light curve of the longer period of both Seq.414 and Seq.232 exhibits an eclipsing behavior, thus the high FAP does not preclude the reality of these two periods. 
In Appendix \ref{sec:append1},  using simulated light curves, we further demonstrate that even a moderate periodic variation can be readily detected by the GL algorithm against strong red noise. 

The remaining three sources with a high FAP, do not exhibit an eclipse that otherwise ascertain the reality of the detected period. 
However, an inspection of the inter-observation light curve of Src-No.481 clearly identifies a peak-to-peak periodic variation (Figure~\ref{fig:erosita_src}), which is captured by both the GL algorithm and the LS periodogram.
Similarly, the inter-observation light curves of Seq.273 and Seq.312 clearly reveal periodic variations  
(see Figure~\ref{fig:longgap}).
Therefore, the detected period in these three sources are probably genuine despite their substantial red noise and the relatively high FAP resulted. 
Nevertheless, we caution that a periodic candidate reported by the GL algorithm with a high, simulation-based FAP could be a spurious detection caused by substantial aperoidic noise.
Moreover, considering the large number of sources searched for periodic signals, a substantial possibility cannot be ruled out that some of the periodic signals with a moderate FAP are also spurious. Additional information about their reality (e.g., eclipsing behavior or optically-confirmed period), as argued for the above sources, are warranted.

Our approach in the above has focused on the actual observational interval to ensure that the variability among different observations are preserved. This neglects potential red noise leakage from lower frequencies (i.e. on time-scales longer than the length of the observations), which might result in an underestimation of the FAP. 
To assess this potential effect, we expand the simulated time interval for those sources with a long period ($P \gtrsim$ 6 hrs), which are more likely to be affected by red noise leakage.  
For each of these sources, we simulate long light curves from the assumed power spectrum that extends to very low frequencies and then randomly extract a segment to match the actual observational length, based on which we re-evaluate the FAP. 
It turns out that the FAP of these long-period sources does not become significantly higher. This is consistent with the suggestion by \citet{2003MNRAS.345.1271V} that the amount of red noise leakage from low frequencies is usually negligible for a flat power spectrum (with a slopes $\alpha <1.5$), which is the case for most of our long-period sources.

\subsection{Uncertainty of the derived periods}
The GL algorithm provides the probability of periodic variation by integrating odds ratio over the period searching range, identifying the period (frequency) as that with the highest odds ratio. 
However, the frequency-dependent odds ratio itself does not quantitatively reflect the uncertainty of the detected period, which should be related to the statistical noise of a given time series. 
To provide a quantitative estimate of the uncertainty in the derived period, we have generated simulated time series of periodic variations (using both sinusoidal and piecewise functions; see Appendix \ref{sec:append1} and Section 3.3 in \citealp{2020MNRAS.498.3513B}) with an additional red noise, which are fed to the GL algorithm to yield a period detection. The input period, the variation amplitude and the total length are approximately the same as the actual cases. Then, for a given detected period, we take the standard deviation of 100 detections based on the simulated light curves, to characterize the relative uncertainty, $\sigma_{\rm P}$, as presented in Table \ref{tab:srcinfo}.
We caution that this uncertainty should be treated as an approximation rather than a mathematically rigorous error.

\section{Spectral Properties and Source classification}
\label{sec:class}

\subsection{X-ray spectral properties of the periodic sources}
\label{subsec:lc_spec}
Before trying to classify the detected periodic sources, we examine their spectral properties, which can hold important information about their nature.
The source and background spectra are extracted from the aperture described in Section~\ref{subsec:appli}, by using the CIAO tool \emph{specextract}, or \emph{srctool} for the eROSITA source Src-No.481.
Two sources, Seq.232 and Seq.198, are caught by some observations to exhibit a high or bursting state. Thus we extract two spectra for each of these two sources to represent the quiescent and bursting states.
Further, the spectra are adaptively binned over 0.5--8 keV (0.2--5 keV for the eROSITA spectrum of Src-No.481) to achieve a minimum of 20 counts and a signal-to-noise ratio (S/N) greater than 2 per bin.
The resultant spectra are analyzed using XSPEC v12.12.0. 

Anticipating that most, if not all, of the periodic sources are LMXBs, CV or ABs, we adopt a fiducial phenomenological spectral model, which consists of a bremsstrahlung continuum and an unknown line-of-sight absorption (\emph{tbabs} in XSPEC), to fit the spectra.
In principle, the X-ray spectra of CVs and ABs are dominated by a collisionally ionized plasma and thus expected to show metal emission lines \citep{2016ApJ...818..136X}. 
However, essentially all 18 sources exhibit a featureless spectrum, which is understood as due to the moderate spectral S/N and the fact that 47 Tuc has a very low metallicity,  {$[\mathrm{Fe} / \mathrm{H}]=-0.76$} (\citealp{1996AJ....112.1487H}; 2010 ed.).
It turns out that the absorbed bremsstrahlung model provides a reasonable fit to all spectra except for five cases, for which we employ a different spectral model.
Seq.263, Seq.273 and Seq.252 are better modeled by a two-bremsstrahlung model. 
Seq.229, known as a WD accreting from a sub-giant companion \citep{2003ApJ...599.1320K}, is well-fitted by an absorbed cooling flow model ({\it mkcflow} in Xspec; \citealp{1988ASIC..229...53M}) with an intrinsic partial absorption ({\it pcfabs} in Xspec; \citealp{2003ApJ...586L..77M}). 
Seq.217, suggested to be a BH-LMXB \citep{2017MNRAS.467.2199B}, shows the most peculiar spectrum with the highest S/N among all sources. 
Instead of a detailed modelling, we simply adopt the best-fit model from \citet{2017MNRAS.467.2199B}, which consists of an optically-thin plasma ({\it vmekal} in Xspec) and a disk reflection component ({\it pexrav} in Xspec), to describe this spectrum.

\begin{figure*}
\centering
\includegraphics[scale=0.89]{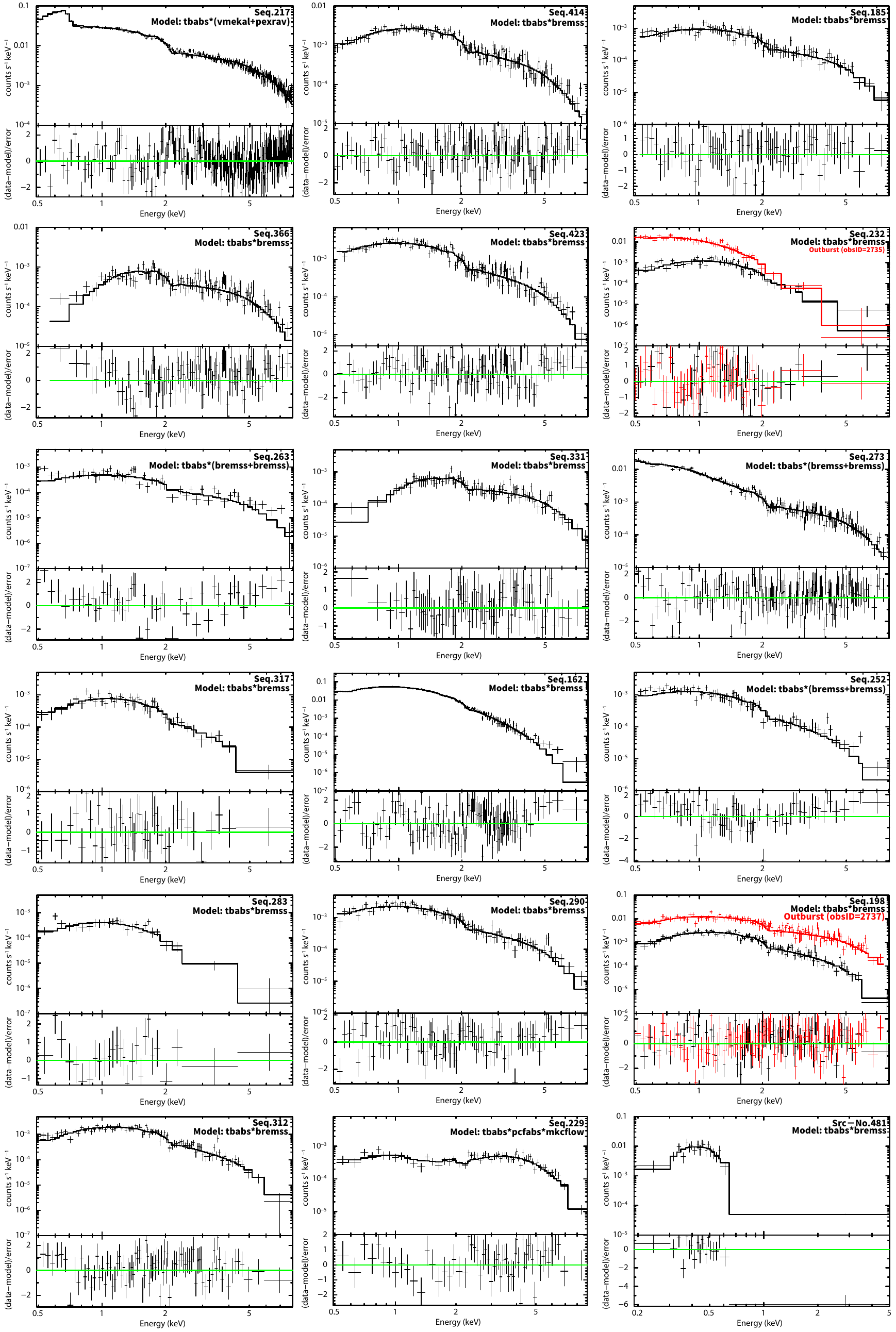}

\caption{Source spectra with the best-fit models. The source sequence number and spectral model are denoted in each panel.
Two sources, Seq.232 and Seq.198, have two sets of spectra respectively from the quiescent (black) and bursting (red) states.} 

\label{fig:spectra}
\end{figure*}

The spectra and best-fit models of the 18 periodic sources are shown in Figure \ref{fig:spectra}. 
Spectral fit results are summarized in Table \ref{tab:spec}, including an unabsorbed 0.5--8 keV luminosity based on the best-fit model and an assumed distance of 4.5 kpc for 47 Tuc (\citealp{1996AJ....112.1487H}; 2010 ed.).

\begin{table}
\centering
\setlength\tabcolsep{3pt} 
\begin{threeparttable}
\caption{X-ray spectral properties of the periodic sources \label{tab:spec}}
\begin{spacing}{1.5}

\begin{tabular}{llllr}
\hline
\hline
Seq & $N_{\rm H}$ & $T_{\rm b}$ & $\chi^2$(d.o.f) & $L_{\rm 0.5-8}$ \\
 & $10^{22}\rm~cm^{-2}$ & keV  & & $\rm 10^{31} erg~s^{-1}$ \\
(1) & (2) & (3) & (4) & (5) \\
\hline

217 & $0.15^{+0.02}_{-0.02}$ & $^a0.17^{+0.01}_{-0.02}$ & 1.30 (229) &  $228^{+16}_{-12}$
\\
414 & $0.10^{+0.03}_{-0.03}$ & $7^{+2}_{-1}$ &  1.04 (133)  & $12.5^{+0.4}_{-0.4}$
\\
185 & 0.023 (fixed) & $9^{+5}_{-3}$ & 0.86 (80)  & $3.7^{+ 0.3}_{-0.3}$
\\
366 & $0.7^{+0.1}_{-0.1}$ & 40 (fixed) &  1.05 (102) & $8.7^{+0.4}_{-0.5}$
\\
423 & $0.04^{+0.03}_{-0.03}$ & $4.6^{+1.2}_{-0.8}$  &  1.11 (113) & $8.6^{+0.3}_{-0.3}$
\\
232 & $0.22^{+0.09}_{-0.08}$ & $0.8^{+0.2}_{-0.2}$ & 1.08 (84) & $3.0 ^{+0.2}_{-0.2}$
\\
232$^\dag$  &  $0.06^{+0.06}_{-0.05}$  & $0.55^{+0.11}_{-0.09}$ & 1.08 (84) &  $14.6 ^{+0.9}_{-0.9}$ 
\\
263 & 0.023 (fixed) & $^b$$0.4^{+0.3}_{-0.2}$(40, fixed) & 1.27 (49) &  $2.1^{+0.2}_{-0.2}$
\\
e481 & $0.08^{+0.08}_{-0.04}$ & $0.06^{+0.01}_{-0.01}$ & 1.00 (16) &  $0.7^{+0.3}_{-0.2}$
\\
331 & $0.7^{+0.2}_{-0.1}$ & $13^{+10}_{-4}$ &  0.70 (78) &  $6.4^{+0.4}_{-0.4}$
\\
273 & 0.023 (fixed) & $^b$$0.26^{+0.02}_{-0.02}$($20^{+25}_{-8}$) & 1.18 (141) &  $21.9^{+0.8}_{-0.8}$
\\
317 & $0.17^{+0.07}_{-0.06}$ & $1.6^{+0.5}_{-0.4}$  & 0.72 (44) &  $2.0^{+0.1}_{-0.1}$
\\
162 & $0.16^{+0.01}_{-0.01}$ & $0.67^{+0.02}_{-0.02}$ & 1.44 (109)  & $102^{+1}_{-1}$
\\
252 & 0.023 (fixed) & $^b$$0.5^{+0.2}_{-0.2}$($19^{+35}_{-9}$) & 0.78 (77) &  $3.9^{+0.2}_{-0.2}$
\\
283 & $0.2^{+0.1}_{-0.1}$ & $0.9^{+0.4}_{-0.3}$ & 1.02 (26) &  $0.93^{+0.09}_{-0.09}$
\\
290 & $0.05^{+0.03}_{-0.03}$ & $4.4^{+1.0}_{-0.4}$ & 0.90 (85)  & $6.4^{+0.3}_{-0.2}$
\\
198 & $0.15^{+0.03}_{-0.03}$ & $2.5^{+0.4}_{-0.3}$ & 1.02 (210)  & $8.5^{+0.4}_{-0.4}$
\\
198$^\dag$ & $0.09^{+0.04}_{-0.03}$ & $19^{+22}_{-8}$ & 1.02 (210)  & $57^{+3}_{-3}$
\\
312 & $0.19^{+0.04}_{-0.04}$ & $2.1^{+0.3}_{-0.3}$ & 0.78 (83) &  $6.1^{+0.3}_{-0.4}$
\\
229 & 0.023 (fixed) & $^c$$11^{+3}_{-2}$ 
	& 1.08 (54) &  $10.4^{+0.6}_{-0.6}$
\\
\hline
\end{tabular}
\end{spacing}
\begin{tablenotes}
      \small
      \item
      Notes: 
      (1) Source sequence number taken from \citet{2019ApJ...876...59C}, except for the eROSITA source e481, which is from \citet{2021arXiv210614535S}.
      $\dag$Spectra at the bursting state.
(2) Line-of-sight absorption column density, fixed at the Galactic foreground value if the spectrum provides no strong constraint.
(3) The bremsstrahlung temperature, fixed at a value of 40 keV if the spectrum provides no strong constraint. 
$^a$Model for this spectrum is not fitted, but is taken from \citet{2017MNRAS.467.2199B};
$^b$Fitted by a two-temperature model; 
$^c$Shock temperature of the cooling flow model. See text for details.
(4) $\chi^2$ and degree of freedom of the best-fit model.
(5) 0.5--8 keV unabsorbed luminosity for a distance of 4.5 kpc, corrected for the enclosed-energy fraction. Quoted errors are at the 90\% confidence level. 
\end{tablenotes} 
\end{threeparttable}
\end{table}

\subsection{Classifying the periodic sources}
\label{subsec:classification}
Now we attempt to classify the periodic X-ray sources, based primarily on their temporal and spectral properties.
In principle, the nature of the periodic X-ray sources can also be inferred from their optical/UV counterparts. However, it is notoriously difficult to distinguish the {\it genuine} counterpart among the numerous GC sources, especially in the dense core.
Nevertheless, we can rely on the periodicity determined by multi-wavelength studies, as a further confirmation of both the X-ray periodic signal and the suggested optical/UV counterpart.
For this purpose, we mainly consult 
\citet{2001ApJ...559.1060A} and \citet{2003ApJ...596.1177E,2003ApJ...596.1197E}, which conducted systematic timing analysis for many tical/UV sources in the core of 47 Tuc and found a number of candidate periods, based on high-resolution, multi-epoch {\it Hubble Space Telescope} (HST) observations. 

Our classification and reasoning for each source are in order below.

{\bf Seq.217 (X9 or W42):} This source was first suggested to be an IP upon the detection of a 218-s periodic modulation in the X-ray light curve \citep{2001Sci...292.2290G}, considered to be a spin modulation.
Here based on much deeper {\it Chandra} observations, we detect a periodic signal at $P = 205.02$ s, with a very small uncertainty ($0.02\%$) that makes it statistically distinct from the 218-s period. 
Unfortunately, \citep{2001Sci...292.2290G} did not provide any detail on how the 218-s signal was determined. Nor could we reproduce this signal using the same observations (in a total exposure of 70 ks) with the GL algorithm.
The phase-folded {at 205 s} light curve shows a sine-like shape with a mild ($\sim 10\%$) fractional variability.
Recently, \citet{2015MNRAS.453.3918M} argued that the high X-ray luminosity ($\rm \gtrsim 10^{33}~erg~s^{-1}$), hard X-ray spectrum and high radio-to-X-ray flux ratio of X9 are inconsistent with a CV, but are more compatible with a BH-LMXB, or possibly a transitional MSP (deemed unlikely from the X-ray spectral and variability properties). A period of $\sim$ 28.18 min and $\sim$ 27.2 min was detected, though at a low significance, by {\it Chandra} X-ray observations \citep{2017MNRAS.467.2199B} and HST far-UV observations \citep{2018MNRAS.476.1889T}, respectively, which were suggested to be the orbital period of the binary system. 
This periodic signal was apparent only in {\it Chandra}/ACIS observations taken in 2002 and only in the 0.5--0.7 keV light curve.
Hence our default period search is not expected to find this signal. 
Nevertheless, when applying the GL algorithm to the 0.5-0.7 keV light curve in the 2002 {\it Chandra}/ACIS observations (ObsID 2735, 2736, 2737, 2738), we are able to recover a periodic signal at $\sim$1692 s with a GL probability of 95\%.
We find it hard to accommodate the dual period with either an IP or a BH-LMXB. In the former case, the longer period of $\sim$1692 s is still too short to be compatible with the orbital period. Moreover, an orbital modulation is unlikely to be transient. 
In the latter case, the longer period of $\sim$1692 s might be due to orbital modulation, but the shorter period ($\sim$205 s) is still too long to be associated with spin modulation of the BH. 
Nevertheless, we follow \citet{2015MNRAS.453.3918M} to tentatively classify this source as an LMXB, chiefly due to the presence of a radio counterpart. The presence of strong carbon lines and lack of H or He lines suggest a CO WD as the donor star \citep{2008ApJ...683.1006K,2015MNRAS.453.3918M}. 
Further studies are warranted to understand its true nature.

{\bf Seq.414 (X13 or W2):} This source was previously reported to have multiple optical periods (2.2, 5.9, and 8.2 hr) and an X-ray period of 6.287 hr \citepalias{2003ApJ...596.1197E}. The latter signal is not recovered in our analysis based on the total {\it Chandra} dataset, nor is it found by the GL algorithm in the single observation studied by \citetalias{2003ApJ...596.1197E}. Hence we consider this signal a false alarm.
Also all the reported  optical periods are suggested to be false alarms by recent work \citep{2018MNRAS.475.4841R}.
We detect two periods with this source, a shorter period at 3846.15 s and a longer period at 8646.78 s.
It is obvious from the phase-folded light curve that the latter arises from an orbital eclipse.
Recent work by \citet{2016MNRAS.462.4371I} also found the same eclipsing period at $\sim 8649$ s from essentially the same {\it Chandra} dataset. Notably, a second, shallower dip is present in the phase-folded light curve, which is separated from the main dip by $\sim$1/2 phase. This raises the possibility that the true period is 50\% less.
On the other hand, \citet{2018MNRAS.475.4841R} found an eclipsing behavior in the optical band at the period of $\sim 8649$ s given by \citet{2016MNRAS.462.4371I}. 
Hence we adopt 8646.78 s as the true period.
The period of 3863.77 s is likely due to spin modulation, which is supported by the sine-like shape of the corresponding phase-folded light curve. 
Therefore, we classify this source as an IP.
The presence of a UV counterpart and the relatively high luminosity $L_{\rm X} \sim 1.2\times 10^{32}\rm~erg~s^{-1}$ are consistent with this classification.

{\bf Seq.185 (W53):} 
The period of 8517.89 s is newly discovered. No periodic signal was found in the optical/UV observations \citepalias{2003ApJ...596.1197E,2018MNRAS.475.4841R}. This source is most likely a CV, indicated by its $L_{\rm X} \sim 3.7 \times 10^{31}\rm~erg~s^{-1}$ and a UV counterpart with a blue color. 
The phase-folded light curve exhibits a wide ($\sim$1/4 phase) but relatively shallow dip, resembling the X-ray (1--9 keV) light curve of the classical polar AM Her \citep{1985A&A...148L..14H}. In this case, the period can also be understood as an orbital period.
The spectrum of this source also appears softer ($T_{\rm b} \sim 9$ keV) than typical IP spectra.

{\bf Seq.366 (W8):} The period of 10311.94 s for this source is newly discovered in the X-ray band. The phase-folded light curve clearly reveals a deep and narrow eclipse, unambiguously pointing to an orbital period. 
The same period and eclipsing behavior were detected in the optical band, but not in the UV band, by \citetalias{2003ApJ...596.1197E}. 
This source is also likely a CV, for its $L_{\rm X} \sim 8.7 \times 10^{31}\rm~erg~s^{-1}$ and UV counterpart with a blue color. 
The plasma temperature ($T_{\rm b}$) is not well constrained for this source, due to its relatively hard X-ray spectrum and insufficient sensitivity of {\it Chandra} at energies above 8 keV.
This suggests that the source is an IP, although no spin modulation is found from the X-ray data. We have fixed $T_{\rm b}$ at 40 keV, typical for IPs when their hard X-ray (up to tens of keV) spectra are available \citep{2016ApJ...818..136X}, to derive $L_{\rm X}$.

{\bf Seq.423 (W1):} 
The period of 10390.33 s is newly discovered in the X-ray band, along with a sub-harmonic at twice this value. These two periods were also detected in the optical band by \citetalias{2003ApJ...596.1177E}, who favored the longer period as the orbital period. The phase-folded X-ray light curve at 10390.33 s exhibits both a dip, likely due to eclipse, and a narrow peak about 1/2 phase ahead of the dip. 
When folded at the sub-harmonic period the two nearly identical dips appear.
Hence we favor the shorter period, i.e., 10390.33 s, as the true orbital period. 
We note that  \citetalias{2003ApJ...596.1197E} favored the longer period not because of a higher significance in the signal itself, but rather due to an empirical relationship between the absolute magnitude of the donor and the orbital period, given that the source is magnetic CV. 

In view of the relatively high $L_{\rm X} \sim 8.6 \times 10^{31}\rm~erg~s^{-1}$ and the UV counterpart, we classify this source as a CV, noting that its soft spectrum ($T_{\rm b} \sim 4.6$ keV) makes it unlikely an IP.

{\bf Seq.232 (W37):} This is another source exhibiting two periods. 
The shorter period, 11112.50 s, is detected only in ObsID 2735, which captured an outburst with an X-ray luminosity ($\sim 1.5 \times 10^{32}\rm~erg~s^{-1}$) a few times higher than the quiescent level.
The same period was reported by \citet{2005ApJ...622..556H}, who suggested W37 to be a quiescent LMXB, although no firm optical counterpart was identified \citepalias{2003ApJ...596.1197E}.
The phase-folded light curve clearly shows a deep eclipse, indicating an orbital period.
A longer period of 44883.30 s is newly detected in the quiescent state, i.e., by excluding ObsID 2735.  Interestingly, this period is consistent with four times the shorter period to within the statistical uncertainty given in Table~\ref{tab:srcinfo}, and the corresponding phase-folded light curve also exhibits an eclipse-like behavior.
One possibility is that the longer period is a super-orbital period, for instance, caused by the precession of a tilted or warped accretion disc. 
We caution that the longer period is more affected by the substantial variability in this source, rendering it a potential spurious signal (Section~\ref{subsec:rednoise}). Further observations are required to confirm this signal and to clarify its relation with the shorter period.
The spectra appear very soft at both the bursting and quiescent states, with $T_{\rm b} \sim$ 0.6 and 0.8 keV, respectively.
This is consistent with thermal X-ray radiation from the surface of an accreting neutron star, as also suggested by \citet{2005ApJ...622..556H}. Hence we also classify this source as an LMXB.

{\bf Seq.263 (W29):} This source is known as
a redback system with a MSP accreting from a main-sequence companion \citep{2002ApJ...579..741E}. Periodic variation at $\sim$ 3.19 hr in the optical and X-rays were found by \citet{2005ApJ...630.1029B}, understood as a changing view of the intra-binary shock between the pulsar wind and the stellar wind from its companion, as described in \citet{2021MNRAS.500.1139H}. 
The radio counterpart, first detected by \citet{2000ApJ...535..975C} using the Parkes radio telescope, with a pulsating period of 2.35 ms and an orbital period of 3.2 hr, exhibits eclipsing-like behavior in the radio light curve for $\sim$25\% of its orbit. 
Here we confirm this orbital behavior at a period of 11486.36 s.
The spectrum is fitted by a two-temperature model, with a low temperature of $\sim$0.4 keV and a high temperature fixed 40 keV, and with one of the lowest X-ray luminosities ($\sim 2.1 \times 10^{31}\rm~erg~s^{-1}$).

{\bf Src-No.481:} This is the only periodic source identified from the eROSITA data and the only one located outside the half-light radius. 
The source was suggested to be a MSP by \citet{2021arXiv210614535S}, based on its soft X-ray spectrum and a potential main-sequence counterpart. 
However, there is no radio detection at the source position in a recent deep radio continuum survey with the Australia Telescope Compact Array \citep{2022MNRAS.513.3818T}, which gives a 5$\sigma$ upper limit of 16 $\rm {\mu}Jy~beam^{-1}$ at 7.25 GHz.
The phase-folded light curve exhibits a single-peaked structure (Figure \ref{fig:erosita_src}), which is consistent with the accretion being dominated by one of the two magnetic poles.
However, the period of 14388.49 s is probably too long for the spin period of a neutron star, but can be compatible with a WD spin.
The latter case, if true, implies that this source is a polar, which often exhibits the one-pole behavior \citep{2020MNRAS.498.3513B}.
On the other hand, the source spectrum is extremely soft, having a best-fit $T_{\rm b} \sim 0.06$ keV, which is more consistent with a MSP \citep{2014ApJ...783...69G,2022MNRAS.511.5964Z} rather than a polar \citep{2016ApJ...818..136X}. 
However, in this case it is hard to reconcile the shape of the phase-folded light curve with orbital modulation.
With the above concerns we tentatively classify this source as a rotation-powered MSP, having a mean X-ray luminosity $\sim 7 \times 10^{30}\rm~erg~s^{-1}$, the lowest among all 18 periodic sources.  

{\bf Seq.331 (W15):} The phase-folded light curve at 15232.63 s clearly shows an eclipse, consistent with previous optical observations \citepalias{2018MNRAS.475.4841R}. With an X-ray luminosity $\sim 6.4 \times 10^{31}\rm~erg~s^{-1}$ and a UV counterpart, this source is likely a CV.
The best-fit $T_{\rm b} \sim 13$ keV is compatible with either magnetic CVs or non-magnetic CVs \citep{2016ApJ...818..136X}.

{\bf Seq.273 (X10 or W27):} 
This source was first suggested to be a CV upon the detection of a $\sim$3.8 hr period in the X-ray band \citep{2001Sci...292.2290G}. Here based on the much deep {\it Chandra} data, we detect a periodic signal at 16824.25 s (4.67 hr), the same value reported before by \citet{2005ApJ...625..796H}. The 3.8-hr signal could not be recovered both in our work and in \citet{2005ApJ...625..796H}.
The phase-folded light curve has a shape reminiscent of the `one-pole' behavior of polars, with the largest fractional variation ($\sim70\%$) among all 18 periodic sources.
The HST observations identified a UV counterpart but found no periodic modulation \citepalias{2018MNRAS.475.4841R}.  
Interestingly, a $\sim$ 18.416 hr periodic modulation has been found in the GeV gamma-ray light curve of 47 Tuc by Fermi observations \citep{2020ApJ...904L..29Z}.
\citet{2020ApJ...904L..29Z} suggested that X10 as a MSP is responsible for the GeV periodic modulation, although Fermi observations lack the angular resolution to directly resolve X10.
However, here we find that the X-ray spectrum of X10 is well characterized by a two-temperature model, with a low-temperature of $\sim$0.26 keV and a high-temperature of $\sim$20 keV. 
This is consistent with typical polar spectra consisting of a soft component from the  blackbody-like WD surface and a hard component from the multi-temperature accretion column \citep{2004MNRAS.347...95R}.
Hence the X-ray spectrum, the phase-folded light curve, the 
X-ray luminosity ($\sim 2.2 \times 10^{32}\rm~erg~s^{-1}$), and the UV counterpart all point to a polar classification for this source.

{\bf Seq.317 (W20):} The period of 23583.11 s is newly discovered in the X-ray band. No periodicity was found by HST observations. 
This source was suggested to be a CV due to the identification of a possible bright optical counterpart  \citepalias{2003ApJ...596.1177E}, while no bright UV counterpart was found that could indicate the presence of an accreting WD \citepalias{2018MNRAS.475.4841R}. The phase-folded light curve, very similar to that of W27, resembles the typical single-peaked light curve of polars. 
On the other hand, its soft X-ray spectrum (with $T_{\rm b} \sim 1.6$ keV) and moderate X-ray luminosity of $\sim 2.0 \times 10^{31}\rm~erg~s^{-1}$ are more suggestive of a DN.
Nevertheless, it seems reasonable to classify this source as a CV.

{\bf Seq.162 (X5 or W58):} We find a period of 31200.27s for this source, which is consistent with the periodic signal reported by \citet{2003ApJ...588..452H} based on early {\it Chandra} observations. The phase-folded light curve clearly shows an eclipse (Figure~\ref{fig:src_example}), thus pointing to an orbital period. 
The soft X-ray spectrum (with $T_{\rm b} \sim 0.67$ keV, but also with significant excess at energies above 5 keV) and a relatively high luminosity ($L_{\rm X}\sim 1 \times 10^{33}\rm~erg~s^{-1}$) suggest that this source is likely an LMXB.

{\bf Seq.252 (W32):}
The source was identified with an optical period of $\sim$ 6.37 hour in \citet{2001ApJ...559.1060A}. Here the X-ray data reveals a new period at 44642.86 s (12.40 hr), consistent with twice the optical period within the statistical uncertainty. 
The phase-folded light curve also appears sine-like.  
A bright, blue, and marginally variable counterpart was suggested by the HST data \citepalias{2003ApJ...596.1197E}.
The X-ray spectrum is well fitted by a two-temperature model, with a low-temperature of $\sim$0.5 keV and a high-temperature of $\sim$19 keV.  
These, together with a moderate X-ray luminosity $\sim 4 \times 10^{31}\rm~erg~s^{-1}$, lead us to tentatively classify the source as a CV.

{\bf Seq.283 (W26):} The X-ray period of 45310.38 s (12.59 hr) is newly discovered. 
A period of 9.45 hr was previously suggested by optical observations, which cannot be recovered by the {\it Chandra} data.
This source is classified as a coronally active binary (AB) based on a possible optical variable counterpart, which has a color consistent with a main-sequence star or a sub-giant \citep{2018MNRAS.475.4841R}.  
The long period, low X-ray luminoisity $\sim 9.3 \times 10^{30}\rm~erg~s^{-1}$, soft spectrum ($T_{\rm b} \sim 0.9$ keV) and the lack of a UV counterpart support an AB classification, although no clear evidence can be inferred from the phase-folded light curve.

{\bf Seq.290 (W25):} The X-ray period of 46082.95 s is newly discovered.
A shallow dip seen in the phase-folded light curve may be attributed to eclipsing at a moderate inclination angle, pointing to an orbital period.
This source was previously classified as a CV but with no optically determined period \citepalias{2003ApJ...596.1177E}. The color of its UV counterpart suggests a hot, relatively low-mass WD \citepalias{2018MNRAS.475.4841R}, and the strong aperiodic variability in the UV supports the CV classification.
The soft spectrum (with $T_{\rm b} \sim 4.4$ keV) and the moderate X-ray luminosity $\sim 6.4 \times 10^{31}\rm~erg~s^{-1}$ are more suggestive of a DN. 
The observation of its NUV transient-like variability, which is likely DN outbursts, provides a strong evidence for its DN nature \citep{2020A&A...634A.132M}.

{\bf Seq.198 (W47 or E8):}
The source has been identified as an AB with an optical period of 12.7 hour, in the absence of blue excess in the HST images \citep{1996ApJ...468..241E,2001Sci...292.2290G}. 
With the deep {\it Chandra} data we confirm the periodic signal at 46151.01 s (12.82 hr), based on the time series excluding the one observation (ObsID 2737) contaminated by flare-like variations. 
The phase-folded light curve is sine-like but with rather moderate fractional variation.
The spectrum of the quiescent state has $T_{\rm b} \sim 2.5$ keV and $L_{\rm X} \sim 8.5\times10^{31}\rm~erg~s^{-1}$, which are not unseen for ABs \citep{2016ApJ...818..136X}.  
The spectrum at the bursting state is much harder and has a luminosity increased by a factor of $\sim7$.
We classify this source as an AB.

{\bf Seq.312 (W23):} 
This source was originally suggested to be an AB exhibiting an optical period of $\sim$ 6.18 hour by \citetalias{2003ApJ...596.1197E}, while recent work \citepalias{2018MNRAS.475.4841R}  classified it as a CV based on a blue-colored UV counterpart and $\rm H\alpha$ emission. 
Here we identify a new periodic signal at 48780.49 s (11.24 hr), with its phase-folded light curve sharing strong resemblance with magnetic CVs, i.e., having a single-peaked structure. Significant variability was found in its UV light curve, by up to 2 mag on a timescale of hours.
Thus we classify this source as a CV, associating the period with spin-orbital modulation. We note that its soft spectrum ($T_{\rm b} \sim 2.1$ keV), moderate luminosity ($\sim 6.1\times10^{31}\rm~erg~s^{-1}$) and phase-folded light curve, similar to that of W20, are more suggestive of a polar.
We note, however, such a period is unusually long compared to the known orbital periods (identical to the spin period) of polars in the solar neighborhood \citep{2003A&A...404..301R}. 
To our knowledge, only one system, V479 Andromedae, having a similarly long orbital period ($\rm P \sim 14.25~hr$), has been suggested to be a polar by its strong He II line in the optical spectrum \citep{2013A&A...553A..28G}.

{\bf Seq.229  (W36 or AKO-9):}  
This source is a spectroscopically confirmed, long-period eclipsing CV system, based on HST observations \citep{1996ApJ...468..241E, 2003ApJ...599.1320K}. 
As mentioned in Section~\ref{subsec:GL}), originally we detect a periodic signal at one-third of the suggested orbital period, mainly due to the limit in our default period-searching range. The true orbital period (95731.20s, or $\sim$26.59 hr) is recovered once the searching range is relaxed. The phase-folded light curve clearly shows the eclipsing behavior.  
The long orbital period requires a subgiant as the companion star, which was confirmed based on HST UV spectroscopy and photometry \citep{2003ApJ...599.1320K}. 
The X-ray spectrum with $T_{\rm b} \sim 11$ keV and $L_{\rm X} \sim 1\times10^{32}\rm~erg~s^{-1}$ are compatible with a DN.

In short, we have classified to a varied degree of confidence five LMXBs, eleven CVs and two ABs (Table~\ref{tab:srcinfo}), based on their X-ray timing and spectral properties and multi-wavelength counterparts. 
In particular, 9 of the 11 CVs have also been identified as CVs in previous work, mainly due to their UV color \citepalias{2018MNRAS.475.4841R}.

\subsection{Remarks on the X-ray-detected periods}

\citetalias{2003ApJ...596.1177E} reported 36 periodic candidates based on multi-epoch HST observations (see table 3 therein, excluding the marginal candidate W120), among which eight were classified as a CV. 
Among them, 5 sources are also  classified as a CV by us (Table~\ref{tab:srcinfo}): W8[=Seq.366], W15[=Seq.331] and W36[=Seq.229] have an X-ray period nearly identical to the optical period; W1[=Seq.423] has an X-ray period one-half of the optical period; W27[=Seq.273] has an X-ray period $\sim20\%$ longer than the optical period. 
For the remaining three periodic CV candidates (W2, W21 and W34) in \citetalias{2003ApJ...596.1177E}, we cannot reproduce the optical period. In particular, as mentioned in Section~\ref{subsec:classification}, W2[=Seq.414] has an X-ray period identical to the optically-confirmed period \citepalias{2018MNRAS.475.4841R}.
The cases of W21 and W34 being a CV have both been  challenged by recent work \citepalias{2018MNRAS.475.4841R}. 

In addition to those CV candidates reported by \citetalias{2003ApJ...596.1177E},
W32[=Seq.252] and W23[=Seq.312] are classified by us as a CV, and their X-ray period are about twice longer than the optical period found by \citet{2001ApJ...559.1060A} and \citetalias{2003ApJ...596.1197E}, respectively.
For the other three CVs classified by us (W53, W20, W25), no periodic signal has been reported from either optical or UV observation. 
Among the non-CV periodic sources reported by \citetalias{2003ApJ...596.1177E},
one source, W29[=Seq.263], is classified as a redback and has a nearly identical period in the X-ray and optical bands;
Two sources are classified as an AB: W47=[Seq.198] has a nearly identical period in the X-ray and optical bands, while W26[=Seq.283] has an X-ray period $\sim30\%$ longer than the optical period. The remaining non-CV periodic sources in \citetalias{2003ApJ...596.1177E} were classified as ABs, but they are either too faint to be detected in the X-ray or with long orbital periods beyond our period-searching range.

To summarize, 6 of the 11 CVs classified in this work show the same period in the X-ray and optical bands, which can be safely taken as the orbital period.
This is particularly true for Seq.414, Seq.366, Seq.423, Seq.331 and Seq.229, the phase-folded light curve of which clearly exhibits an eclipse.  
The X-ray period of the other 5 CVs is not detected in the optical or UV, but can also be taken as the orbital period. Support for this comes from the sign of an eclipse (Seq.185 and Seq.290) or the sign of spin-orbit modulation (Seq.273, Seq.317 and Seq.312).   
One LMXB (Seq.263) and one AB (Seq.198) also exhibit the same period in the X-ray and optical bands, which is most likely the orbital period.
The X-ray period detected in the other four candidate LMXBs is not seen in the optical band.
Among them, two LMXBs (Seq.232 and Seq.162) clearly exhibit an eclipse in the phase-folded light curve, unambiguously pointing to an orbital period. 
On the other hand, the nature of Seq.217 and Src-No.481 remain elusive, so does their X-ray period. 
The X-ray period of Seq.283, another candidate AB, is not seen in the optical, either, but is also tentatively taken as the orbital period.

\section{Discussion}
\label{sec:discussion}
The periodic X-ray sources identified in previous sections can provide an important diagnosis to the underlying close binary populations in 47 Tuc, which in turn allows for a meaningful comparison with the field populations, in particular CVs.  Nevertheless, it is necessary to first examine the purity and completeness of the periodic sources (Section~\ref{subsec:purity}), before implications on their demography and origin can be addressed (Section~\ref{subsec:LW}).

\subsection{Purity and completeness of the periodic X-ray sources} 
\label{subsec:purity}
It is possible that some of the periodic X-ray sources are background AGNs, which are known to exhibit in their X-ray emission quasi-periodic oscillations (QPOs) as well as the more dramatic quasi-periodic eruptions (QPEs) in the frequency range of $10^{-4}-10^{-2}$ Hz \citep{2016ApJ...819L..19P,2018ApJ...853..193Z,2018MNRAS.477.3178C,2019Natur.573..381M,2020A&A...644L...9S,2021Natur.592..704A}.
Recently, \citet{2022MNRAS.509.3504B} conducted a systematic search for AGN QPOs from the 7-Ms {\it Chandra} Deep Field-South (CDF-S; \citealp{2017ApJS..228....2L}), which consists of $\sim 1000$ independent X-ray sources. 
Using both the LS periodogram and the GL algorithm, \citet{2022MNRAS.509.3504B} found no statistically significant periodic signals that persist over the entire 7-Ms exposures, or over any of the four subsets having an effective exposure of 1-Ms to 3-Ms. 
From the empirical log$N$--log$S$ distribution of cosmic X-ray background (CXB), \citet{2019ApJ...876...59C} estimated the number of background AGNs to be $\lesssim 4.8$ within a projected radius of $1^\prime$ from the center of 47 Tuc, i.e., within which all the detected periodic sources are located. 
Even the much larger eROSITA FoV, when compromised by its moderate sensitivity, is expected to contain $\lesssim$ 200 background AGNs.
A simple scaling from the non-detection of QPOs among the $\sim1000$ CDF-S sources suggests that the possibility of having an AGN QPO in the field of 47 Tuc is negligibly small.
The moderately high Galactic latitude of 47 Tuc ($b \approx -44\fdg9$) also implies that contamination by foreground sources is negligible. 
\citet{2019ApJ...876...59C} also found negligible contamination by the Small Magellanic Cloud, which is an extended background object near the line-of-sight toward 47 Tuc.
Thus it is safe to conclude that all 18 periodic X-ray sources are physically associated with 47 Tuc.

Completeness of the periodic sources is less straightforward to assess. 
For this purpose, we contrast with the work of \citep{2020MNRAS.498.3513B}, who found 23 periodic sources among 847 X-ray sources in the Limiting Window (LW), based on 1-Ms {\it Chandra} exposure of the Limiting Window (LW) and the GL algorithm. 
The point-source sensitivity limit of the LW reaches $L_{\rm X} \sim 10^{30} \rm~erg~s^{-1}$ \citep{2010A&A...513A..63R}, which is only slightly highly than that of 47 Tuc ($L_{\rm X} \sim 4 \times 10^{29} \rm~erg~s^{-1}$). Moreover, the detected periodic sources typically have a flux significantly above the sensitivity limit. Thus it is statistically meaningful to compare the two samples.
\citet{2020MNRAS.498.3513B} assessed the completeness of periodic signal detection with the GL algorithm by simulating a large set of X-ray light curves, which have the characteristic shape of either a sine function or an eclipse and cover the ranges of orbital period, fractional variation amplitude and source counts typical of the LW sources (mostly CVs and ABs). 
They found that for a sinusoidal light curve, the GL detection completeness is generally $\gtrsim20\%$ for source counts $C \gtrsim 100$ and exceeds $\sim$50\% for $C \gtrsim 300$; the detection completeness rapidly rises to $\gtrsim 90\%$ at high variation amplitudes ($\gtrsim 60\%$).
For an eclipsing light curve, the GL detection completeness is lower for the same number of counts and exceeds 50\% only with $C \gtrsim 800$. In both cases, the detection completeness increases with increasing period. 
These simulation results can be borrowed to evaluate the present case, since the 47 Tuc sources have quite similar ranges of periods and source counts, but also with an on-average lower background.
We estimate that $C \sim 100$, corresponding to $L_{\rm X} \sim 3 \times 10^{30} \rm~erg~s^{-1}$ in 47 Tuc, can be taken as the detection limit for a typical periodic signal, whereas $C \gtrsim 300$, corresponding to $L_{\rm X} \gtrsim 10^{31} \rm erg~s^{-1}$, can be regarded as the completeness limit. 
This is consistent with the lowest luminosity of $9 \times 10^{30} \rm~erg~s^{-1}$ found among the periodic sources (excluding the eROSITA source Src-No.481).

\begin{figure*}
\centering
\includegraphics[scale=0.56]{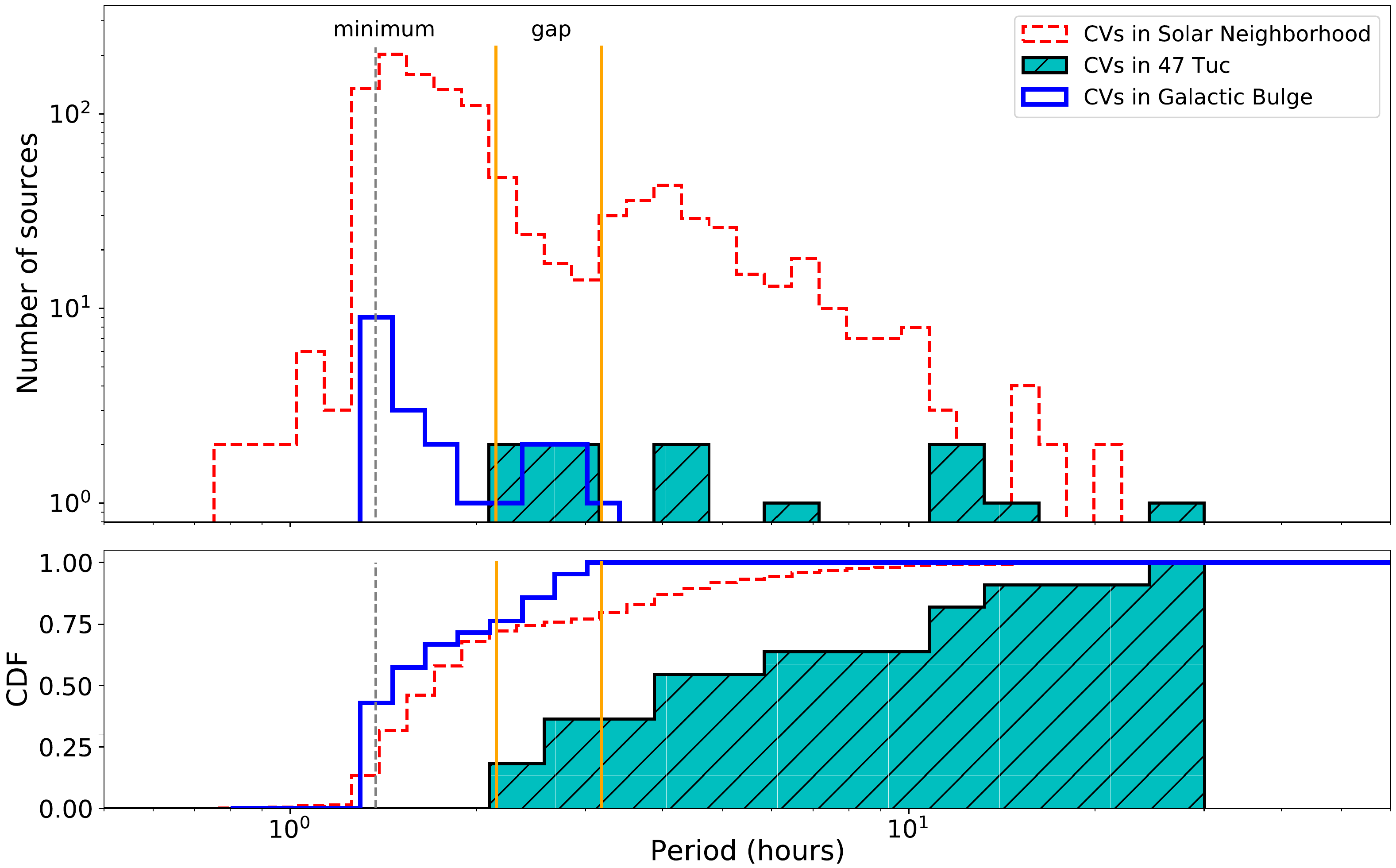}
\caption{The orbital period distribution of the 47 Tuc CVs (cyan-filled histogram), in comparison with that of field CVs (red-dotted histpgram)and galactic bulge CVs (blue histpgram). 
Period gap and period minimum are delineated by a pair of vertical orange-solid lines, and a vertical grey-dashed line, respectively, the values of which are taken from \citet{2011ApJS..194...28K}. The bottom panel shows the cumulative distribution.}
\label{fig:N_P}
\end{figure*}

\begin{figure*}
\centering
\includegraphics[scale=0.54]{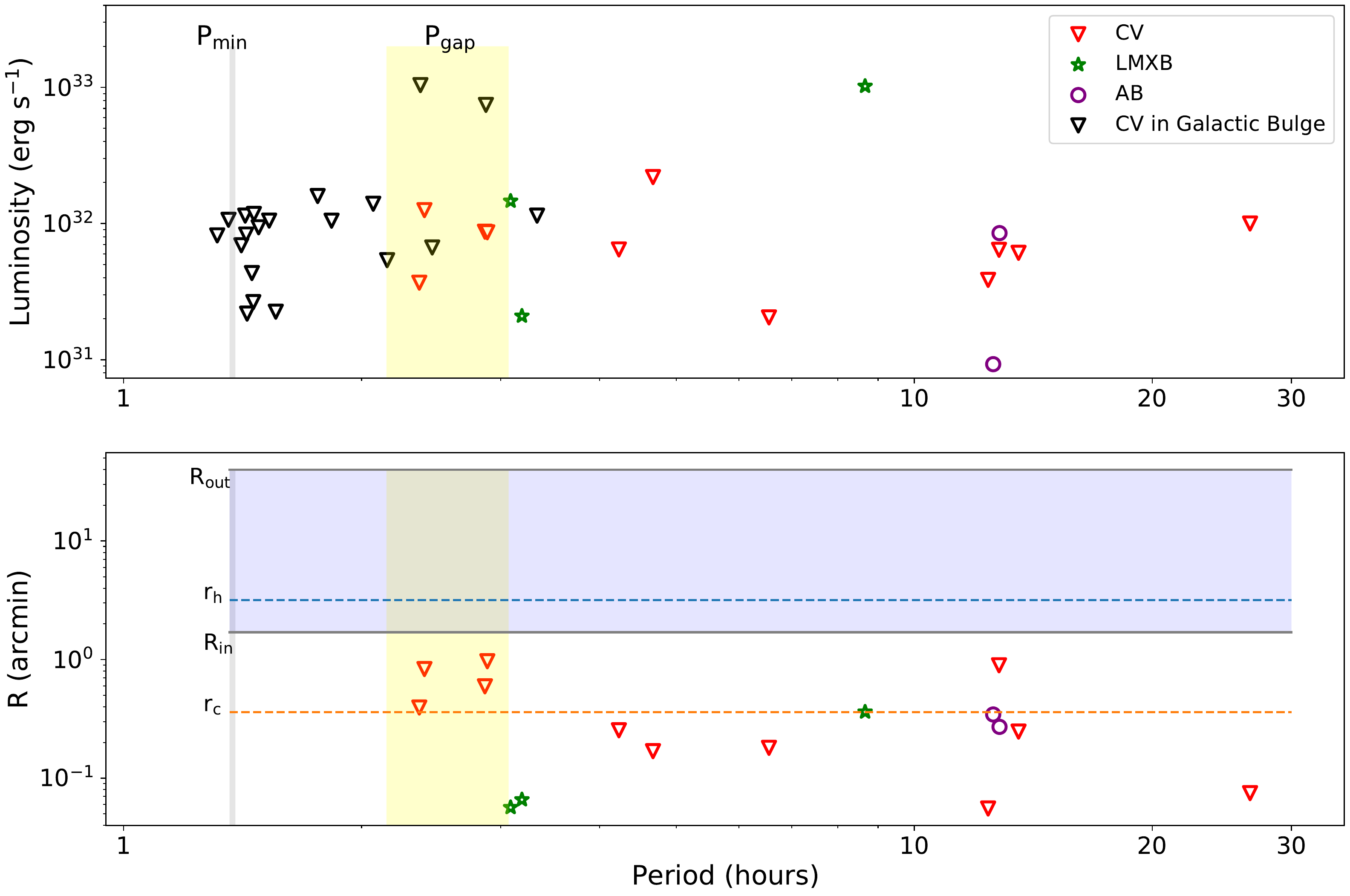}
\caption{{\it Upper panel}: The 0.5--8 keV X-ray Luminosity versus orbital period, for CVs (red triangles), ABs (purple circles) and LMXBs (green stars) identified in 47 Tuc. Galactic bulge CVs (black triangles) with an identified orbital period are plotted for comparison. 
{\it Low panel}: The project distance from the center of 47 Tuc versus orbital period. 
The core radius ($r_{\rm c}$) and half-light radius ($r_{\rm h}$) are denoted by the orange and blue dashed line, respectively. 
The horizontal blue-shaded area marks the eROSITA field-of-view, which extends to the cluster's tidal radius. 
In both panels, the period gap and period minimum of CVs, with values taken from \citet{2011ApJS..194...28K}, are denoted by the yellow strip and the grey vertical line, respectively.}
\label{fig:profile}
\end{figure*}

\subsection{Orbital period distribution and origin of CVs}
\label{subsec:LW}

Now we turn to the orbital period distribution of the X-ray sources, focusing on the CV candidates. 
Figure \ref{fig:N_P} displays the histogram of the putative orbital period of the 11 CVs in 47 Tuc. For comparison, we also show the orbital period distributions of CVs in the solar neighborhood from \citet{2003A&A...404..301R} and twenty CVs in the Galactic bulge (i.e., the LW) from \citet{2020MNRAS.498.3513B}. 
The sample of solar neighborhood CVs (red dashed histogram) has a much larger size than the bulge (blue solid histogram) and 47 Tuc (green filled and hatched histogram) samples, which clearly exhibits the two well-known features, namely, the period gap between $\sim$ 2--3 hr and the period minimum at $\sim$82 min, as marked in Figure \ref{fig:N_P}.
Notably, the Galactic bulge CVs are almost all found within or below the period gap, having a longest period of 3.33 hr. According to \citet{2020MNRAS.498.3513B}, this is not due to selection bias but can be understood as an age effect, in the sense that the Galactic bulge CVs are more evolved systems, having substantially shrunk orbits compared to their solar neighborhood counterparts.  

Several interesting aspects of the 47 Tuc sample are revealed in this diagram:

(i) No CV is found to have a period below 2.15 hr, the lower bound of the period gap \citep{2011ApJS..194...28K}. This is in stark contrast with the bulge and solar neighborhood CVs, which have $75\pm25\%$ (15 of 20) and $69\pm3\%$ (769/1116) found below the gap, respectively. Here the quoted errors are of 1-$\sigma$ Poisson errors. 

(ii) Long period CVs, i.e., those having a period beyond the upper bound of the gap (3.18 hr), contribute to $64\pm30\%$ (7/11) of the 47 Tuc sample. This is  significantly higher than the fraction of $5\pm4\%$ (1 of 20) and $23\pm2\%$ (254 of 1116) in the bulge and solar neighborhood sample, respectively. Moreover, 4 of the 11 CVs in 47 Tuc have an orbital period above 12 hr, which is rarely seen even in the much larger sample of solar neighborhood CVs ($\sim 1\%$, 13 of 1116).

(iii) $36\pm20\%$ (4 of 11) of the 47 Tuc sample is found with the gap, while this fraction is $20\pm10\%$ (4 of 20) and only $8\pm1\%$ (93 of 1116) in the bulge and solar neighborhood sample, respectively. 

Figure \ref{fig:profile} further depicts the CV orbital period versus the X-ray luminosity (upper panel) and the projected radius from the center of 47 Tuc (lower panel). For completeness, sources classified as LMXBs and ABs are included, except for Seq.217 (X9) and Src-No.481, the X-ray period of which is of an unclear origin (Section~\ref{subsec:classification}). 
For Seq.232 and Seq.198, the luminosity at the quiescent state is plotted.
The Galactic bulge CVs with an identified orbital period \citep{2020MNRAS.498.3513B} are also included in the upper panel for comparison.
The 0.5--8 keV luminosities of these sources have been derived from the best-fit spectral model in \citet{2020MNRAS.498.3513B}. 
In the lower panel, we use a blue-shaded area to represent the effective eROSITA FoV, with an inner radius of $R_{\rm in} = 1\farcm7$ and an outer radius of $R_{\rm our} = 42\arcmin$. The latter is coincident with the cluster's tidal radius \citep{1996AJ....112.1487H}.
The half-light radius ($r_{\rm h} = 3\farcm17$) and core radius ($r_{\rm c} = 0\farcm36$) of 47 Tuc are also denoted. From Figure \ref{fig:profile}, an additional aspect of the 47 Tuc CVs can be stated:

(iv) More than half (6 of 11) of the CVs with an identified period are located within the core radius, which include six of the seven long-period CVs. On the other hand, all four CVs within the period gap are located outside the core radius. 
In the meantime, no CVs with an identified period are found outside $R = 1\arcmin$, or about one-third of the half-light radius. 

The above outlined behavior of the CV orbital periods have important implications about the origin of present-day CVs in 47 Tuc. 
For (i), the lack of short-period CVs appears contradictory to the theoretical expectation that short-period CVs, especially period bouncers, should dominate the present-day CV population in GCs \citep{2016MNRAS.462.2950B}, which essentially reflects the fact that the evolutionary timescale of CVs is short compared to the age of GCs. 
We emphasize that this paucity of short-period CVs is not solely a selection effect, since the simulations presented in \citet{2020MNRAS.498.3513B} indicate that the current {\it Chandra} data are sensitive to periodic signals between 1--2 hr and of an X-ray luminosity down to $\sim10^{31}{\rm~erg~s^{-1}}$. Indeed, most LW CVs with an identified orbital period are found below the gap and have $L_{\rm X} \gtrsim 2\times10^{31}{\rm~erg~s^{-1}}$.
Thus the paucity of short-period CVs in 47 Tuc suggests a CV population different from those in the Galactic bulge and solar neighborhood. 
We note that the majority of the 20 LW CVs were classified as a magnetic CV (IP or polar), mainly based on the ``one-peaked'' behavior in their phase-folded light curve \citep{2020MNRAS.498.3513B}. 
\citet{2020MNRAS.498.3513B} further demonstrated that the detectability of orbital modulation for DNe, typically through an eclipse, is generally much lower than the detectability for magnetic CVs. 
This is consistent with the analysis of field CVs using ASCA observations, which found that 11 of 16 IPs exhibit an X-ray orbital modulation \citep{2005A&A...439..213P}, while only 4 of 34 DNe do so \citep{2005MNRAS.357..626B}. 
This implies that the non-detection of short-period CVs might be attributed to an intrinsically lower fraction of magnetic CVs among all CVs in 47 Tuc, compared to the field population, in which a fraction of 23\% \citep{2003A&A...404..301R} to 36\% \citep{2020MNRAS.494.3799P} has been estimated. 

On the other hand, period bouncers as low-luminosity systems are unlikely to be detected with the current {\it Chandra} data.
This can be inferred from the flux-limited X-ray surveys conducted by \citet{2012MNRAS.419.1442P}, which suggest that period bouncers would have $L_{\rm X} \lesssim \rm 10^{30}~erg~s^{-1}$, according to the empirical luminosity distribution of non-magnetic CVs.
CVs with such low X-ray luminosities may still be detected as an X-ray source, but the small number of counts ($\lesssim$ a few tens) they produce are insufficient for the GL algorithm to reveal their orbital period.

For (ii), the high fraction of long-period CVs in 47 Tuc is also at odds with the GC age. 
The low fraction of long-period CVs in the Galactic bulge compared to the field has been explained by an age effect \citep{2020MNRAS.498.3513B}.
Given the age of GCs, the intuitive expectation is that 47 Tuc would have a similar or even lower fraction of long-period CVs compared to the bulge, which, however, is not observed.
We emphasize that three of the four 47 Tuc CVs with an identified period above 12 hr have a bright, variable UV counterpart, arguing strongly for their CV identity. While only one of them (Seq.229) shows a clear sign of orbital eclipse, such long periods cannot be attributed to the WD spin period and are most probably the orbital period.
Given the wide orbits, the donor should be a sub-giant in order to fill the Roche lobe. The progenitor mass of the donors in these long period CVs should be higher than the main-sequence turnoff mass, $\sim 0.9~\rm M_{\odot}$, of 47 Tuc \citep{1987PASP...99..739H, 2009AJ....138.1455B} and in turn requires a more massive WD.
This strongly suggests a dynamical origin for these systems (e.g., via exchange or tidal capture), which is consistent with the fact that three of the four systems are located within the core radius (Figure~\ref{fig:profile}), where the stellar dynamical encounter rate is highest. 
A possible formation channel is predicted by the simulations of \citet{2006ApJ...646..464S} (see model CV1 in Table 2 therein), in which weak orbital perturbations force the mass transfer start when the donor is still on the subgiant branch, having $P_{\rm orb} \sim 11.5~\rm hr$. The CV phase of this case lasts for $\sim$ 16 Myr, until $P_{\rm orb}$ reduces to $\sim 6.5~\rm hr$, allowing for the onset of the common envelope (CE) phase. 
Generally, the exchange interaction is most likely to produce CVs with donor mass greater than 0.7 $\rm M_{\odot}$, and the mass-transfer phase is short-lived, on the order of 10--100 Myr. 
This means that these long-period CVs formed only recently. 

The fact that these long-period ($P_{\rm orb} > 10$ hr) CVs have only moderate X-ray luminosities ($L_{\rm X} \lesssim 10^{32}{\rm~erg~s^{-1}}$) are also remarkable. 
In fact they are no more luminous than their short-period counterparts, whereas standard evolution models of CVs generally predict that the mass transfer rate is more than one order of magnitude higher in long-period CVs \citep{2016ApJ...833...83K}, which should lead to a higher $L_{\rm X}$.
In view of the combined properties of a long orbital period and an evolved donor, these sources are analogous to several field CVs including 1RXH J082623.6-505741 ($P_{\rm orb} \sim $ 10.4 hr, \citealp{2022arXiv220610625S}), EY Cyg ($P_{\rm orb} \sim $ 11 hr, \citealp{2007A&A...462.1069E,2020AdSpR..66.1139N}), CXOGBS J175553.2-281633 ($P_{\rm orb} \sim $ 10.3 hr,  \citealp{2021MNRAS.502...48G}) and KIC5608384 ($P_{\rm orb} \sim $ 9 hr,  \citealp{2019MNRAS.489.1023Y}). The low mass transfer rates inferred for these binaries are still not well understood. For example, the observed mass transfer rate of KIC5608384, is a factor of 20 lower than predicted by a MESA model \citep{2019MNRAS.489.1023Y}. 
The presence of long-period CVs with relatively low X-ray luminosities in a massive GC invites further investigation.

As for (iii), finding a significant fraction of CVs within the period gap is even more unexpected. Theoretically, non-magnetic CVs are not able to sustain mass transfer within the period gap, since the contraction of the donor on a thermal timescale is faster than the orbital shrinking on the much longer gravitational radiation timescale, making the system detached \citep{1988A&A...202...93R}. 
Two possible explanations may be considered for these systems. The first possibility is that they are magnetic CVs, thus their mass transfer is controlled by the magnetic field of the WD and remains substantial across the gap \citep{2018ApJ...868...60G}. 
We note that this could be the case for the four LW CVs found within the gap (Figure~\ref{fig:profile}).
In 47 Tuc, a magnetic CV is highly likely the case for Seq.414, which exhibits a dual-period and is best understood as an IP (Section~\ref{subsec:classification}). The hard X-ray spectrum of Seq.366 is also suggestive of an IP, while the light curve of Seq.185 is suggestive of a polar (Section~\ref{subsec:classification}). 
The problem with this explanation is that we should expect to detect other magnetic CVs below the gap, in view that short-period CVs make up 48\% of all magnetic CVs in the solar neighborhood \citep{2003A&A...404..301R}. However, none is observed, which, as we argue in the above, may be understood as due to a low occupation fraction of magnetic CVs.    
This leads us to consider the second possibility, that is, these systems are dynamically formed (e.g., via exchange), or their progenitor binaries have been dynamically altered, such that they are semi-detached when crossing the period gap. 
We remind that these four CVs are all currently located outside the core radius. Therefore, if they indeed formed via dynamical encounters, the encounter most likely took place inside the core and must have provided the resultant binary with a sufficiently large radial velocity to escape from the core.
Finding such CVs outside the core is reasonable, since the two-body relaxation time at the half-light radius of 47 Tuc ($T_{\rm rel} \sim 3.5$ Gyr, \citealp{2019MNRAS.483..315B}) is much longer than the CV evolutionary time crossing the gap ($\sim 0.4-1.3$ Gyr) \citep{2011ApJS..194...28K}. 

\begin{figure}
\centering
\includegraphics[scale=0.46]{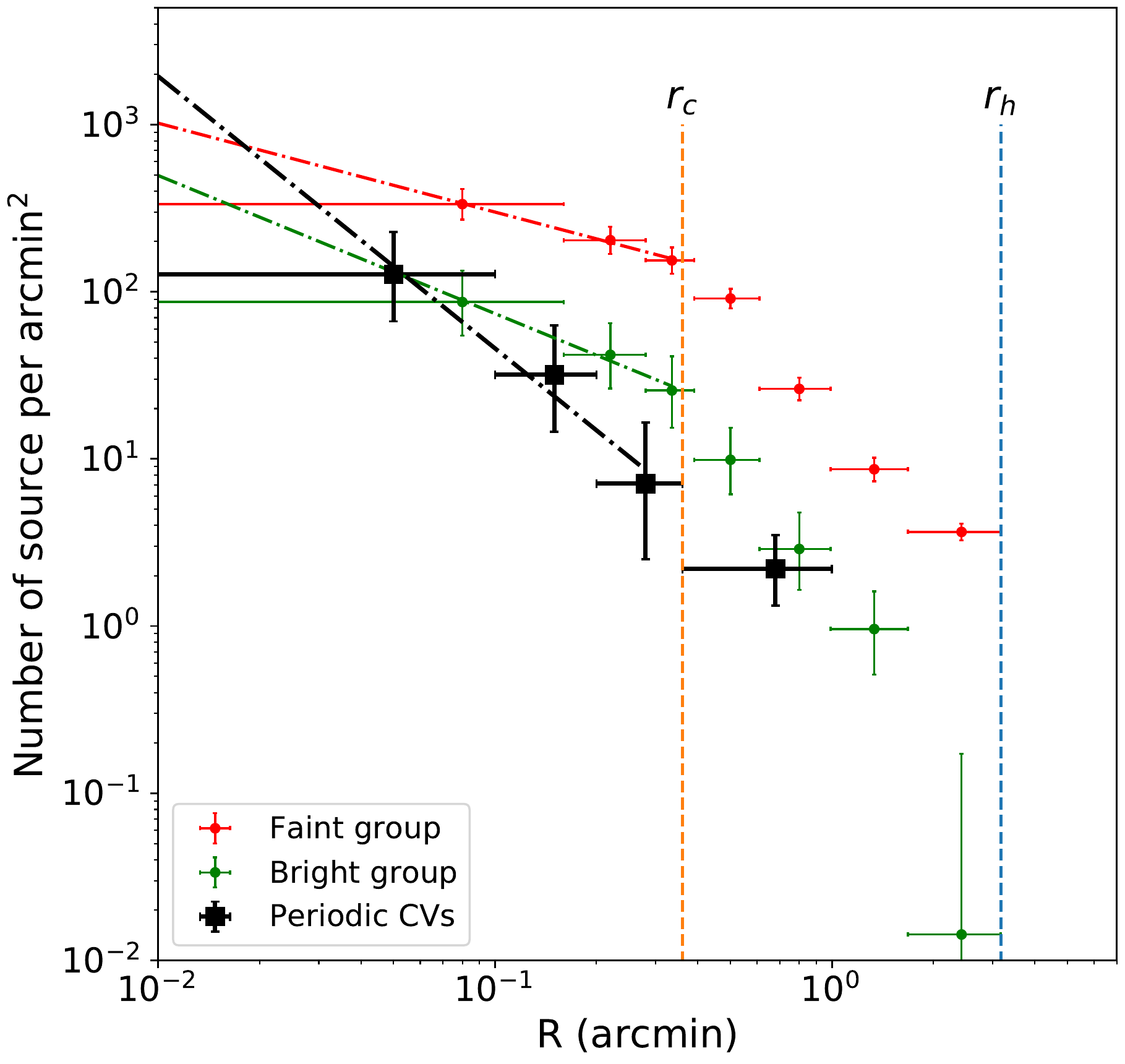}

\caption{Radial surface density profiles of the periodic CVs (black), bright X-ray sources ($L_{\rm X} \gtrsim  5 \times 10^{30}{\rm~erg~s^{-1}}$, green), faint X-ray sources ($L_{\rm X} \lesssim  5 \times 10^{30}{\rm~erg~s^{-1}}$, red). The CXB contribution has been subtracted. 
The colored dash-dotted lines represent the best-fit power-law function of three profiles out to the core radius.
The half-light radius and core radius are mark by the blue and orange lines.}

\label{fig:profile_3g}
 	
 \end{figure}

For (iv), as already discussed in the above, the locations of the CVs provide a strong hint for their possible dynamical origin. 
We further construct the radial surface density profile for the 11 periodic X-ray sources classified as a CV, as shown in Figure~\ref{fig:profile_3g} (black squares). 
For comparison, we also plot the radial surface density profiles of all the {\it Chandra}-detected X-ray sources within the half-light radius of 47 Tuc, excluding the periodic sources.
Following \citet{2019ApJ...876...59C}, these sources are divided into two groups: the bright group 
($L_{\rm X} \gtrsim 5 \times 10^{30}{\rm~erg~s^{-1}}$, green dots) and the
faint group ($L_{\rm X} \lesssim 5 \times 10^{30}{\rm~erg~s^{-1}}$, red dots). 
Contamination by the CXB sources has also been statistically subtracted (see Figure 5 in \citealp{2019ApJ...876...59C} for detail).
\citet{2019ApJ...876...59C} showed that both the bright and faint groups are more centrally concentrated than the normal stars in 47 Tuc, which can be understood as the effect of mass-segregation, i.e., the X-ray sources as close binaries are on-average heavier than the single stars and are more likely to sink into the cluster core through two-body relaxation.  
Remarkably, Figure~\ref{fig:profile_3g} shows that the periodic CVs have an even steeper radial distribution than either the bright group or the faint group, which is most prominent within the core radius. 
To demonstrate this more quantitatively, we fit the three radial profiles out to the core radius using a power-law function, finding a slope of $-1.6\pm0.3$, $-0.8\pm0.1$ and $-0.53\pm0.04$ (1$\sigma$ errors) for the periodic CVs, the bright group and the faint group, respectively. 
Since the bright group should mainly consist of CVs and quiescent LMXBs, the effect of mass-segregation alone is unlikely to explain the steeper profile of the periodic sources. 
We suggest that the latter reflect a subgroup of CVs having recently formed out of two-body or multi-body interactions inside the cluster core, which would naturally follow a steep radial profile. 

The lack of periodic sources outside one-third of the half-radius, expect for the unclassified source Src-No.481, deserves some remarks.
We note that this apparent lack is not due to an insufficient sensitivity. 
First, the {\it Chandra} data has a nearly constant detection sensitivity out to $r_{\rm h}$ \citep{2019ApJ...876...59C}, meaning that any periodic source having similar properties with those detected should be found by the GL algorithm with an equal probability. However, none of the 156 sources located between 1/3 $r_{\rm h}$ to 1 $r_{\rm h}$ exhibits a significant periodic signal.
Second, the eROSITA observations have a source detection sensitivity of $5 \times10^{30}\rm~erg~s^{-1}$ out to a radius of at least 18\farcm8 \citep{2021arXiv210614535S}. 
Although ideally one would use simulations to determine the precise sensitivity of the eROSITA data for periodic signal detection, we may empirically estimate that this sensitivity limit for periodic signals is about one order of magnitude higher than the source detection limit, i.e., $5 \times10^{31}\rm~erg~s^{-1}$.
We note that the luminosity of Src-No.481 is significantly below this limit, due to its unusual softness, but otherwise supports our conservative estimate. 
After accounting for the CXB contribution, we further estimate that there are 
$\sim$ 80--90 sources ($\sim$ 20 have $L_{\rm X} > 5 \times10^{31}\rm~erg~s^{-1}$), out of a total of 237 eROSITA sources lying between $R = 1\farcm7$ and 18\farcm8, which should belong to 47 Tuc.  
Therefore, even with a 10\% GL detection efficiency on average, we should expect to find 8--9 periodic sources, while only one is actually observed.
This, along with the {\it Chandra} non-detection of periodic sources outside $R = 1\arcmin$, suggests an intrinsic paucity of bright CVs at large radii.
Since bright CVs and their progenitor binaries should be substantially heavier than the background stars, 
they could have sunk to the cluster core. 
This is in accord with the abrupt drop of the surface density profile of the bright group at $R \gtrsim 2\arcmin$ (Figure~\ref{fig:profile_3g}), or a ``dip'' identified by \citet{2019ApJ...876...59C}.
On the other hand, CVs recently formed in the cluster core via dynamical interactions may have received a substantial kick velocity that could take them to the cluster outskirt.  
Indeed, the MOCCA simulation of \citet{2019MNRAS.483..315B} predicts that on average about half of the entire population of detectable CVs are currently located outside the half-light radius. 
This discrepancy between the observation and simulation deserves further study.

\section{Summary}
\label{sec:summary}
We have presented a systematic search for X-ray periodic signals among 537 sources resolved by {\it Chandra} and 888 sources resolved by eROSITA in the field of 47 Tuc. 

Our main results include:

\begin{itemize}

\item Using the Gregory-Loredo algorithm, we have detected 20 X-ray periodic signals from 18 independent sources, among which 17 are detected by {\it Chandra} and one by eROSITA. Fourteen of the 20 periodic signals are newly discovered. 

\item We have tentatively classified the 18 periodic sources into 11 CVs, 4 LMXBs, 1 MSP and 2 ABs, based primarily on their X-ray temporal and spectral properties, as well as optical/UV information, when available. 

\item Despite a sample subject to potential selection bias and small-number statistics, the orbital period distribution defined by the 11 CVs in 47 Tuc is significantly different from that of the Galactic bulge and solar neighborhood CVs. The paucity of short-period CVs below the period gap might be attributed to a high occupation fraction of non-magnetic CVs in 47 Tuc, for which the orbital modulation is more difficult to detect by the GL algorithm.
In the meantime, both the overabundance of long-period CVs with a subgiant donor and a substantial fraction of CVs found within the period gap provide strong evidence for a group of CVs having recently formed via dynamical interactions.

\item The steep radial surface density profile of the periodic CVs provides further evidence for a group of CVs having recently formed out of dynamical interactions inside the cluster core. In the meantime, despite sufficient sensitivity of the X-ray data, no periodic CVs are detected outside one-third of the half-light radius and up to the cluster's tidal radius. 

In future work, we plan to extend the search for periodic X-ray sources to other Galactic GCs, which will be valuable for a better understanding of the periodic features found in 47 Tuc and, more generally, in the dynamical environment of GCs.

\end{itemize}

\section*{Acknowledgements}
This work is supported by the National Natural Science Foundation of China (grant 12225302) and the science research grants from the China Manned Spaced Project (CMS-CSST-2021-B02, CMS-CSST-2021-B11). B.T. acknowledges support by the Postgraduate Research \& Practice Innovation Program of Jiangsu Province (KYCX22\_0106). 
 Z.C. acknowledges support by the National Natural Science Foundation of China (grant 12003017).
The authors wish to thank the anonymous referee for helpful comments that improve our work, Ziteng Wang for valuable discussions, and Lulu Xing for help with the graphics.
\section*{Data Availability}
The data underlying this article will be shared on reasonable request to the corresponding author.



\bibliography{pXS_Tuc}
\bibliographystyle{mnras}




\appendix
\section{Additional consideration on red noise}
\label{sec:append1}
As discussed in Section \ref{subsec:rednoise}, 
several periodic sources are accompanied with substantial red noise, which may give rise to a relatively high false detection probability. 
However, the presence of red noise does not preclude the detection of true periodic variation. 
This can be demonstrated as follows. 
 
Specifically, we consider the possibility of detecting a periodic signal, which assumes an intrinsic sinusoidal form, in the presence of red noise. 
The periodic variation is defined as $f(t)=A \times \lambda sin(\omega t +\phi)$, where $\lambda$ represents the mean source count rate, $\omega = 2\pi/P$ is the frequency, and $A$ is the variation amplitude. 
We adopt $A=0.5$ and $P$=48780.49 s, i.e., nearly identical to the actual case of Seq.312, as trials. The phase $\phi$ is fixed at zero since it has no effect on the detection rate.  
The red noise is based on the best-fit power spectral distribution for Seq.312 (Figure~\ref{fig:312psd}).
We follow the method described in Section \ref{subsec:rednoise} to generate a group of 1000 time series, combing the red noise and the sinusoidal variation. 
These time series are then fed to the GL algorithm over the searching range of (10000, 50000) s.
It turns out that a periodic signal is detected in all 1000 simulated time series with a reported period within 0.1\% of the input period and a GL probability greater than 0.99.
This exercise thus demonstrates the efficiency of the GL algorithm in detecting periodic signals against substantial red noise. 

\section{Inter-observation light curves of Seq.312 and Seq.273 showing periodic variations}
\label{sec:append2}
\begin{figure}
\centering

\includegraphics[scale=0.33]{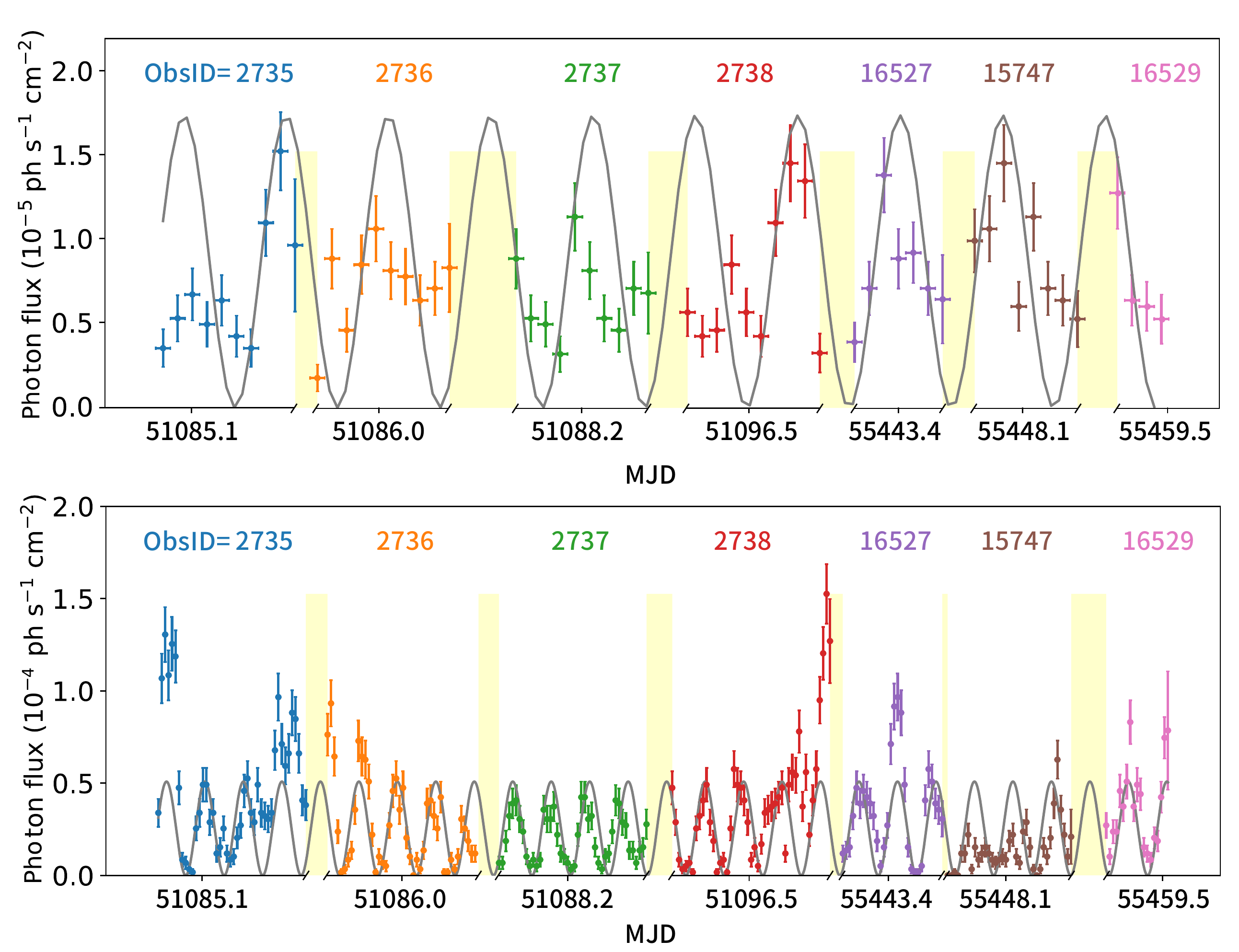}

\caption{The {\it Chandra} 0.5--8 keV long-term light curve of Seq.312 (upper panel) and Seq.273 (lower panel), covering several long observations that together reveal the periodic variation (with a sinusoidal curve overlaid to guide the eye). 
Each set of colored data points are from a single ObsID (as labelled) and have a bin size of 7000 s for Seq.312 and 1500 s for Seq.273.
The yellow strips mark the eliminated gap (an integer multiplication of the period)  between two consecutive observations. 
} 
\label{fig:longgap}
\end{figure}

The long-term light curve of the {\it eROSITA} source Src-No.481, as shown in Figure \ref{fig:erosita_src}, exhibits quite regular periodic variation. Thanks to the very small observation gaps, the periodicity is appreciable to the naked eye in this nearly continuous light curve. 
We perform a similar exercise for Seq.312 and Seq.273.
Despite the more irregular cadence of the {\it Chandra} observations, 
we manage to construct the inter-observation light curves from several long observations, after eliminating the observation gaps with an integer multiplication of the detected period. 
As shown in Figure \ref{fig:longgap}, despite substantial aperiodic variability in both sources, their inter-observation light curve exhibits a peak-to-peak variation following the detected period, as in the case of Src-No.481.
This strongly suggests that the detected periods are intrinsic, rather than a fake signal caused by the red noise.

\end{document}